\documentstyle[aps,epsfig,preprint,tighten,float]{revtex}  
\newcommand{\be}{\begin{equation}}
\newcommand{\ee}{\end{equation}}

\begin{document}
\title{Ground State and Spin Glass Phase of the Large $N$ Infinite Range Spin 
Glass Via Supersymmetry}
\author{M. B. Hastings}
\address{Physics Department, Jadwin Hall\\
Princeton, NJ 08544\\
hastings@feynman.princeton.edu}
\maketitle
\begin{abstract}
The large $N$ infinite range spin glass is considered, in particular the
number of spin components $k$ needed to form the ground state and the 
sample-to-sample fluctuations in the Lagrange multiplier field on each site.
The physical significance of $k$ for the correlation functions is discussed.
The difference between the large $N$ and spherical spin glass is emphasized;
a slight difference between the average Lagrange multiplier of the large $N$ 
and spherical spin glasses is derived, leading to a slight increase
in the energy of the ground state compared to the naive expectation.
Further, there is a change in the low energy density of excitations in
the large $N$ system.
A form of level repulsion, similar to that found in random matrix
theory, is found to exist in this system, surviving interactions.
Even though the system is an interacting one, a supersymmetric formalism
is developed to deal with the problem of averaging over disorder.
   \end{abstract}
\section{Introduction}
Spin glasses have been extensively investigated for a long 
time\cite{classi1,classi2}.  
Spin glasses with finite $N$ (the definition of
$N$ is discussed below) are dealt with using
a replica technique which works wonderfully in the infinite range model.
Unfortunately, it is known from other problems
in random matrix theory that\cite{susyrmt} that replica techniques sometimes
fail, especially for nonperturbative properties.  
In this paper we look at the large $N$ spin glass, and show that there
exists a supersymmetric technique for dealing with this system.
This system is a disordered, interacting (although interacting only in
the sense of the large $N$ limit) system, that can still be dealt with in
a fairly rigorous fashion.

The large $N$ generalization of the spin glass that we use is not new, although
we will show
that the behavior is much more complicated than has been found using
replica techniques on this problem\cite{oldstuff}.  For example,
we define a number $k$, the number of spin components used to form the
ground state, and find a non-trivial scaling of $k$ with the 
system size; in
the replica technique $k$ is not even defined.  We will find significant
differences between the large $N$ spin glass and finite $N$ spin glass,
much greater than the differences between large $N$ and finite $N$ unfrustrated
systems.  Many of
these differences stem from the fact that while an unfrustrated system
will have the same ground state for large $N$ as for finite $N$, the frustrated
large $N$ systems will use a large number, $k$, of the available spin
components; the finite $N$ systems have far fewer spin components
available.

Of course, all physical glassy systems will have $N=1,2,$ or $3$,
and will not be large $N$ systems. The
different behavior of the large $N$ system in the glassy case
makes the large $N$ approximation less useful for glassy
systems than it is for ordered systems.  However, 
there may exist some optimization problems, or problems from computer
science or combinatorics, which will lie in the large $N$ limit, at least
for finite $V$.  
Also, it may be possible someday to extend the techniques in this
paper to the finite dimensional case, and obtain results on this system
which cannot be obtained in the finite $N$, finite dimensional problem.

The infinite range model mentioned above is one of the simplest models of a 
spin glass.  It is a system of $V$ spins, all interacting with each either via 
Gaussian distributed
random interactions.  In the general case, we may let each spin have up to
$N$ components, so that the Hamiltonian is given by
\be
\label{hdef}
H=\sum\limits_{i,j,\mu}\phi_i^{\mu} \phi_j^{\mu} H_{ij}
\ee
where $i,j$ index the various sites and range from $1$ to $V$, while $\mu$
indexes various components of the spin and ranges from $1$ to $N$.  We
impose the constraint that
\be
\label{normal}
\sum\limits_{\mu}(\phi_i^{\mu})^2=N
\ee
We will take $H_{ij}$ to be a real symmetric matrix for most of the paper.
In section XV we will discuss situations in which $H$ may have
complex entries and $\phi$ may also be complex.

We will consider the problem in the large $N$ limit.  Consider the
statistical mechanics of the system, integrating over all spin configurations
with a weight $e^{-\beta H}$.  It is well-known, and
will be reviewed below, that in the large $N$ limit
the solution of a self-consistency equation will provide the 
properties of the system at any given inverse temperature $\beta$
and for any given matrix $H_{ij}$.
A similar self-consistency equation will provide the ground state properties
of the system.

The problem that will be considered in this paper is the ground
state and thermodynamic properties of the Hamiltonian given by
equations (\ref{hdef},\ref{normal}), in the large $N$ limit, when $H_{ij}$ is 
drawn from an ensemble of real, symmetric, Gaussian distributed matrices.  
There exist several solutions of such problems in the 
literature\cite{oldstuff}, using replica techniques.
However, there are several problems with such solutions.  Later in the
paper we will discuss technical problems with that work.  However, here,
let us simply mention that two important properties of the ground state
(and the spin glass phase) are not addressed in that work.  One property
is the number of spin components used to form the ground state.  Another is
the strength of sample-to-sample fluctuations in the Lagrange multiplier
field used to define the large $N$ limit (discussed in a later section).

Consider the number of components used to form the ground state.
It is clear that a system with large $N$ has more freedom to choose a
ground state than a system with a smaller $N$.  There are more possible
directions to point the spin components.  To give a simple example, consider
a system with three sites, and an anti-ferromagnetic interaction between the
sites, so that $H_{ij}=1$ for all $i,j$.  If $N=1$, the lowest energy state
is to take one site with spin $+1$ and two sites with spin $-1$, or vice-versa.
If $N=2$, one can find a lower energy state.  For example, one can set
$\phi_1=\sqrt{2}(1,0)$, $\phi_2=\sqrt{2}(-1/2,\sqrt{3}/2)$, and 
$\phi_3=\sqrt{2}(-1/2,-\sqrt{3/2})$.  The factor of $\sqrt{2}$ in front
of all the spins is simply to conform with the normalization chosen in
equation (\ref{normal}); the important point is that the spins now
use two different components.  For the same system, with $N>2$, one will
still find that the spins only need to use two components.  In fact, for
this system, for any $N$, the different spins will always span a
2 dimensional vector space.

Let us define the number of spins used to form a given state.  We
will call this number $k$.  We may define this number for any state as
the dimension of the vector space spanned by the $V$ different
vectors $\phi_i^{\mu}$, where here as $i=1...V$ the spins
$\phi_i^{\mu}$ define $V$ different vectors 
lying in the $N$ dimensional space of vectors $\phi^{\mu}$.  The vector space 
spanned by these vectors is a $k$-dimensional subspace of this $N$-dimensional 
space.

The ground state of a system will have Goldstone modes, due to the
possibility of rotating all the spins together.  Ignoring these modes,
a system will generically have a non-degenerate ground state; that is, all 
ground states will be linked by symmetry.  In that case, the number of spins 
needed to form the ground state will
be the dimension $k$ of the space spanned by the vectors $\phi_i^{\mu}$
used to form the ground state.  However, we can imagine systems with degenerate
ground states.  For example, a system of $V$ spins, with no interaction
between the spins, has a very degenerate ground state manifold.  All
states have the same energy, so all states are ground states.  The value
of $k$ for these states can be anything between $1$ and $V$.  When we
refer to the number of spins needed to form the ground state, we will
in this case use the smallest $k$ of all the states on the ground
state manifold.

An important question is how big $k$ is for the ground state in the 
large $N$ limit, where the system can take $k$ as large as it wants.  One 
property we will find, due to the additional freedom to place spins in more 
directions for large $N$, is the uniqueness of the solution to the 
thermodynamic self-consistency equation, and the absence of metastable 
states.  This may be interpreted as an absence of replica symmetry breaking.

The central result of the paper is the relation between $V$ and the
number of spin components $k$ used to form the ground state.  We will show
that $k(k+1)/2\leq V$ for any given $H_{ij}$.  Then we will show
that the infinite range system system, for large $V$, has $k$ of order
$V^{2/5}$.  So, the bound is not saturated.

Throughout the paper, we will be taking the large $N$ limit first, and then
considering systems with large, but finite $V$.  By addressing the number of
spin components used to given $V$, we will then be able to comment on the
nature of the large $N$ and large $V$ limits.  In fact, these limits
are different, and the interchange of limits is not justified.   One
can guess that if $N$ is of order $V^{2/5}$, or bigger, then the system
will be in a regime described by large $N$, while if $N$ is smaller than
$V^{2/5}$ the behavior will be completely different.  It is likely that
replica symmetry breaking will occur only if $N$ is smaller than $V^{2/5}$.

From results on random matrices\cite{rmt}, the idea of level repulsion
is common.  We will show in section II that for a given $k$, there 
will be a matrix which has $k$ eigenvalues
all of order $1/\beta$, so that in the large $\beta$ limit these eigenvalues
collide.  This might seem to contradict the naive expectations of level
repulsion; we will show in section V that a modified form of level repulsion 
still exists in the system.

Another important result obtained in this paper
is the magnitude of the sample-to-sample and
site-to-site fluctuations in the Lagrange multiplier field $\lambda_i$
used to define the large $N$ limit.  This field also has a physical
significance, connected with the total energy of the bonds attached 
to a given site.  This will provide information on the sample-to-sample
variations in total energy.  Within the replica formulations used
previously, there are problems with calculating these fluctuations; this
and other problems with the replica formulation will be discussed later.

From a standpoint of formal technique, it is nice that we can use
supersymmetric techniques.  This is a glassy, interacting system, and
it is useful to have such a system that can be treated without
using replicas.  Although the replica technique is
very powerful, it sometimes has certain mathematical difficulties 
that a supersymmetric technique does not.  For calculation of the
spectra and correlators of random matrices, the Efetov supersymmetry techniques
are superior\cite{susyrmt}.  The supersymmetric technique used here
is more closely connected to the supersymmetry introduced by Parisi
and Sourlas\cite{parisour}.  However, we will make some connections
to the Efetov supersymmetry technique in section XIII.

The paper is set up as follows: first, we introduce the problem.  In
section II we review the solution of the large $N$ problem via
self-consistency.  For any given $H$, we can exactly solve the problem by
solving a self-consistency equation.  The problem is to obtain results
after averaging over different $H$.  In section III and section IV we
review theorems\cite{me} showing that the solution of the self-consistency
equation is unique and bounding the value of $k$.  

After this introduction and review of old results, we discuss
some of what should be expected from the solution of the large $N$
spin glass.  We discuss level repulsion, paramagnetic and spin glass phases, 
and correlation functions.
Section V introduces
the idea of level repulsion in the interacting system.  Section VI
discusses the spherical model and section VII discusses the large $N$
model in the paramagnetic phase.  In the paramagnetic phase, these
two models are equivalent in the thermodynamic limit.  In the spin
glass phase they are very different, and this difference is emphasized.
In section VII we also discuss an analogy between $k$ and the existence
of Griffiths effects in finite dimensional systems.
Section VIII discusses the consequences of the quantity $k$ for the
correlation functions and the physical significance of $k$.  

Then we proceed to the actual calculation for the large $N$ system.  In section
IX the supersymmetric formalism is derived.  In section X a simple
example is discussed to help introduce some of the manipulations
used in section XI when the supersymmetric formalism is used to
obtain results on the ground state.  In section XII we discuss the
physical meaning of these results from section XI, and argue that
the results of that section are also valid for the spin glass phase.  

Finally, we make some connections with Efetov supersymmetry and replica
techniques, and discuss extensions of the problem.  Section XIII shows
some interesting connections between the supersymmetric formalism of
this paper and the Efetov supersymmetry.  Section XIV reviews problems
with the replica solutions of the problem.  Section XV discusses extensions
of the results in this paper to more general matrices $H$, and section XVI
is a conclusion.  At the very end there are two appendices 
containing some of the
mathematical manipulations needed for the calculations of section XI.

Finally, we will include a note on notation.  Throughout, we will
refer to vectors $\phi_i^{\mu}$ or $v_i^{\mu}$.  
The index $i$ will range from $1...V$
and will index different sites in the problem.  The index $\mu$ will 
range either from $1...k$ or from $1...V$ and can be
used to index different spin components on each site, or to index one of
$k$ or $V$ different eigenvectors (these vectors lie in the $V$-dimensional
space of vectors $\phi_i$).   Eigenvalues will be written as $E_{\mu}$.
Lagrange multiplier fields which are a function of the site will be
written as $\lambda_i, c_i,$ and so on.
\section{Large $N$ and Self-consistency}
In this section, we will review the solution of the large $N$ problem
via a self-consistency equation.  First we will discuss the solution of
the thermodynamics of the problem at finite $\beta$.  Then we will discuss
the self-consistency equation for the ground state.

For the thermodynamics of the problem, we wish to compute the following
partition function
\be
Z=\int\limits_{\sum\limits_{\mu} (\phi_i^{\mu})^2=N} d\phi_i^{\mu} e^{-\beta H}
\ee
where
\be
H=\sum\limits_{i,j,\mu}\phi_i^{\mu} \phi_j^{\mu} H_{ij}
\ee
Rewriting the constraint on the spins using a Lagrange multiplier field
$\lambda_i$, we can write the partition function as
\be
\int d\phi_i^{\mu} \, d\lambda_i e^{-\phi_i 
(\beta H_{ij} +\lambda_i\delta_{ij})
\phi_j} e^{N\lambda_i}
\ee
where the integral over $\lambda_i$ extends from $-i\infty$ to $+i\infty$.
It is convenient to rescale $\lambda_i$ by a factor $\beta$ to obtain
\be
\int d\phi_i^{\mu} \, d\lambda_i e^{-\phi_i 
(\beta (H_{ij} +\lambda_i\delta_{ij}))
\phi_j} e^{N\beta\lambda_i}
\ee

The next step is to integrate out the field $\phi$, to derive an action
for the Lagrange multiplier field.  Then, after finding a saddle point for
this field, it may be verified in the large $N$ limit that fluctuations about
the saddle point are small.

The partition function after integrating out $\phi$ is
\be
\int d\lambda_i e^{-N{\rm Tr \, ln}
 (\beta (H_{ij} +\lambda_i\delta_{ij}))}
 e^{N\beta\lambda_i}
\ee
The saddle point equation for $\lambda_i$ is then
\be
\label{selcon2}
G_{ii}=\beta
\ee
where we define the Green's function matrix $G$ by
\be
(H_{ij}+\lambda_{i})^{-1}=G
\ee
In the large $N$ limit, after solving for $\lambda_i$, then the field $\phi$
can be treated as a free Gaussian field, with Hamiltonian
\be
\beta (H_{ij}+\lambda_{i})
\ee
When solving for $\lambda_i$, we must find a $\lambda_i$ such that
\be
\label{poscon}
H_{ij}+\lambda_{i} > 0
\ee
That is, $H+\lambda_i$ must be a positive definite matrix.

The average energy of the system is given by the derivative of the
log of the partition function with respect to $\beta$.  The result is
\be
\label{energyeq}
\frac{V}{\beta}-\sum\limits_{i=1}^{V} \lambda_i
\ee

We can take a zero temperature ($\beta\rightarrow\infty$) limit of the 
self-consistency equation (\ref{selcon2})
to find an equation defining the ground state of the system.  
Since equation (\ref{selcon2}) implies that $G_{ii}$ must diverge in
this limit, some number of eigenvalues of $H+\lambda_i$ must go to zero.

In the zero temperature limit of the problem, we can also solve
the problem by giving a configuration of spins, $\phi_i^{\mu}$ which minimizes
the energy.  It is clear that if the given configuration is a local
minimum of the energy, then for some $\lambda_i$
\be
\label{zersc}
(H_{ij}+\lambda_{i}\delta_{ij})\phi_j^{\mu}=0
\ee
for all $\mu$.  That is, the spin configuration must be made up of zero
eigenvectors of $H_{ij}+\lambda_{i}\delta_{ij}$.  If the spins span a
$k$ dimensional space, then there must be at least $k$ such zero eigenvectors.

In fact, we can write the zero temperature problem as a problem of
finding $\phi_i^{\mu}$ and $\lambda_i$ such that equation (\ref{zersc})
is satisfied, such that all spins have  length equal to $\sqrt{N}$, and 
such that 
\be
H_{ij}+\lambda_{i}\delta_{ij}\geq 0
\ee
That is, the matrix $H_{ij}+\lambda_i\delta_{ij}$ must be positive
semi-definite.

To give a simple example in which $k \neq 1$, consider the system of
three spins interacting anti-ferromagnetically with each other, as
discussed in the introduction.  We have $H_{ij}=1$ for all $i \neq j$.
If we take $\lambda_i=1$ for all $i$, we find that $H_{ij}+\lambda_i$
has two zero eigenvectors.  We can choose to write these as 
$\sqrt{\frac{2}{3}}(1,-1/2,-1/2)$ and 
$\sqrt{\frac{2}{3}}(0,\sqrt{3}/2,-\sqrt{3}/2)$.  By taking
$\phi_1=\sqrt{2}(1,0)$, $\phi_2=\sqrt{2}(-1/2,\sqrt{3}/2)$, and 
$\phi_3=\sqrt{2}(-1/2,-\sqrt{3/2})$, we have expressed the spins $\phi_i^{\mu}$
as linear combinations of these eigenvectors, and satisfied the constraint
on the length of the spin on each site.  Here we have a system with $k=2$.

For large, but finite $\beta$, the matrix $H+\lambda_i$ will have
$k$ eigenvalues which scale to zero as $1/\beta$.  These are the eigenvalues
which will vanish in the zero temperature limit.  The other $V-k$
eigenvalues will tend to non-zero limits in the limit of large $\beta$.
If these eigenvalues are denoted $E_{\mu}$ for $\mu=1...k$, with eigenvectors
$v_i^{\mu}$, then an acceptable ground state is
\be
\label{normali}
\phi_i^{\mu}=\sqrt{\frac{N}{\beta E_{\mu}}}v_i^{\mu}
\ee

In the thermodynamic limit, it is possible for there to be a phase
transition as the temperature is lowered.  In this case, $H+\lambda_i$ will have
$k$ eigenvalues of order $k/V$.  As a simple example, consider a system
of $V$ spins, such that $H_{ij}=-1/V$ for all $i,j$.  
This is a system with a ferromagnetic
interaction between all the spins.  For any $\beta$, the self-consistency
equation will be solved by taking $\lambda_i$ to be the same for all $i$.
Then, $H+\lambda_i$ has $V-1$ eigenvalues equal to $\lambda$ and one
eigenvalue equal to $\lambda-1$.  The self-consistency equation
becomes
\be
\label{example}
\frac{V-1}{V}\lambda^{-1}+\frac{1}{V}(\lambda-1)^{-1}=\beta
\ee
For large $V$, this is solved in the high temperature, small $\beta$,
region by taking $\lambda=\beta^{-1}$.  As $\beta$ increases, however, there
occurs a phase transition at $\beta=1$.  Past the phase transition,
we must always have $\lambda>1$, in order for $H+\lambda$ to
be positive definite.  In fact, $\lambda$ will be equal to $1+{\rm O}(1/V)$.
The self-consistency equation (\ref{example}) can be approximated by
\be
1+\frac{1}{V}(\lambda-1)^{-1}=\beta
\ee
We see that $V-1$ eigenvalues remain gapped in the large $V$ limit, while
one eigenvalue becomes of order $1/V$.  Physically speaking, this is
a state which is macroscopically occupied.  For a frustrated system, in which
$k$ eigenvalues become macroscopically occupied, there will be $k$ eigenvalues
of order $k/V$.

We expect that the properties of the system on the ordered side of
the phase transition will be similar to the zero temperature properties.
At zero temperature, $k$ eigenvalues becomes equal to zero, vanishing
as $1/\beta$.  For finite,
but large, $\beta$, we have $k$ eigenvalues of order $k/V$.  For large $V$,
the eigenvalues are all close to zero.  So, it is reasonable to expect
that the properties of the spin glass phase should be similar to the
zero temperature problem.

In the simple example of a ferromagnet, we had a gap between the one
eigenvalue which was macroscopically occupied and the $V-1$ eigenvalues
above it.  This will not remain true in the spin glass systems considered
in this paper.  There will be a spectrum of excitations above the
$k$ eigenvalues which are macroscopically occupied.  This spectrum
will have a small gap which vanishes as a power law in the large $V$
limit; we will derive this gap in section XI and appendix A.  All
the states above the gap can be treated as a continuum and integrated
over when obtaining the Green's function; there will be $k$ states below
the gap which must be treated more carefully.
\section{Uniqueness of Self-consistent Solution}
Given the self-consistency equation defined in the previous section, we
will show that the solution of this equation is unique.  That is, we
have the equation for all $i$
\be
G_{ii}=\beta
\ee
where we define the Green's function matrix $G$ by
\be
(H_{ij}+\lambda_{i})^{-1}=G
\ee
We will show that for any $H$ there is only one solution $\lambda_i$
of this equation which has the property that $H_{ij}+\lambda_i$ is
a positive definite operator.

The proof of this has been given before\cite{me}, but we will review it
again here.  The proof proceeds as follows: assume for some given $H$, we
have a $\lambda_i$ that solves the self-consistency equation, with 
$H+\lambda_i$
positive definite.  We will show below that for any infinitesimal 
change ${\rm d} H$ in $H$, we can find a change ${\rm d} \lambda_i$ 
in $\lambda_i$ 
such that the self-consistency equation is still obeyed.  Also, we will show 
that the change
in $\lambda_i$ is well-behaved; that is, that the differential equation we
will find leads to $\frac{{\rm d}\lambda_i}{{\rm d}H_{ij}}$ always
being finite.
These two results imply that starting from any given $H$, we can deform $H$ 
from the 
initial $H$ to the point $H=0$, and correspondingly deform $\lambda_i$ along 
this path, with the deformation of $\lambda_i$ along the path being
unique and well-behaved.  
However, for $H=0$, the only solution is clearly $\lambda_i=1$ 
for all $i$.  Now, if the initial $H$ were to have two different 
solutions to the self-consistency equation with different $\lambda_i$, then 
there would be a contradiction, since then we could find two different 
solutions to the self-consistency equation for $H=0$ by deforming
the two different solutions from the initial $H$ to $H=0$.  So, the theorem 
will be proved.

We will now demonstrate that $\frac{{\rm d}\lambda_i}{{\rm d}H_{ij}}$ 
always is finite, as stated above.  Afterwards, we will show that the
deformations of $H$ and $\lambda_i$ always keep $H+\lambda_i$ positive definite.

Consider the equation
\be
G_{ii}=\beta
\ee
Consider changes ${\rm d}H$ and ${\rm d}\lambda_i$.  Then we will have
\be
G_{ii}+\sum\limits_{j,k}G_{ij}({\rm d}H_{jk}+{\rm d}\lambda_j \delta_{jk})G_{ki}
=\beta
\ee
Then we have
\be
\sum\limits_{j}G_{ij} {\rm d}\lambda_j G_{ji}
=-\sum\limits_{j,k}G_{ij}{\rm d}H_{jk}G_{ki}
\ee
The right-hand side of this equation is some function of $i$.  The left-hand
side is some symmetric
linear operator, acting on $\lambda_j$.  To prove the desired
result we need to show that this linear operator is always invertible.
This will follow if it can be shown that the linear operator is positive
definite.  To show that, we simply need to show that
\be
\label{inteq}
\sum\limits_{i,j}{\rm d}\lambda_i G_{ij} {\rm d}\lambda_j G_{ji}>0
\ee
for all vectors ${\rm d}\lambda_i$.  However, equation (\ref{inteq})
is an immediate consequence of $G$ being a positive definite operator.  

Now we will show that under these deformations $G$
remains positive definite, which is equivalent to $H+\lambda$
remaining positive definite.   It is impossible for $G$ to acquire a zero 
eigenvalue
since this would require that $H+\lambda_i$ have an infinite eigenvalue which
cannot happen while $H$ and $\lambda_i$ remain finite.  
Then, the only way for $G$ to acquire a negative eigenvalue is
for one of the eigenvalues of $H+\lambda_i$ to pass through zero.  However,
this would require that one of the eigenvalues of $G$ diverge.  Looking at
the self-consistency equation (\ref{selcon2}), and taking a trace, we find that
\be
\label{sowesee}
Tr(G)=\beta V
\ee
where $V$ is the number of sites.  Since all eigenvalues of $G$ are positive,
equation (\ref{sowesee}) implies that any given eigenvalue is 
bounded by $V$, so it is not possible for an
eigenvalue to diverge.

This concludes the proof.  We have shown that, starting from a given $H$,
and a solution of the self-consistency equation, it is always possible to
deform to $H=0$, while staying in the sector in which $H+\lambda_i$ is
positive definite, and with the change in $\lambda_i$ always a regular function
of the change in $H$.
\section{Maximal Number of Components to Form the Ground State}
We will derive a bound on the maximal number of components needed to form
the ground state of an arbitrary Hamiltonian for a system of $V$ spins. 
The system will be in the large $N$ limit so that it has as many components
available as it needs.  This theorem has been given before\cite{me}, but
it will be reviewed here.  We will consider here only the case of
real Hamiltonians; others will be discussed in the last section.
The result for real Hamiltonians is that $k$, the number of components needed, 
satisfies the inequality $k(k+1)/2\leq V$.  At the end of this section
we will explicitly construct a sequence of systems which saturate this
bound for any $k$.

The proof of the inequality $k(k+1)/2\leq V$ is very simple.
Suppose the system needs $k$ spin components to form the ground state.
Then, for small perturbations $\delta H$ of the Hamiltonian $H$, the system
will continue to need $k$ spin components to form the ground state.
However, as shown above in the discussion of the large $N$ self-consistency
equation, this means that the operator $H+\lambda_i$ has at least $k$
zero eigenvalues.  For a real, symmetric matrix, such as $H+\lambda_i$, to
have $k$ eigenvalues requires adjusting a total of $k(k+1)/2$ parameters.
Since we have assumed that for any small perturbation of $H$ the system
continues to have $k$ zero eigenvalues, the only parameters available
to tune to maintain the $k$ zero eigenvalues are the $V$ different values
of $\lambda_i$.  Therefore, we have that $k(k+1)/2\leq V$.

Note that this result is not a result for an ensemble of $H$; it is
a result for any $H$.  To express the proof more formally, let $H$ be
a $V$-by-$V$ matrix.  The dimension of the space of $\delta H$ is
$V(V+1)/2$.  The space of matrices $(H+\delta H +\lambda_i +\delta \lambda_i)$
that satisfy equation (\ref{selcon2}) has dimension $V(V+1)/2-V$ since 
we have $V$ parameters $\delta \lambda_i$
to adjust to satisfy the $V$ self-consistency equations.  However, the
space of $V$-by-$V$ matrices with $k$ zero eigenvalues has dimension $V(V+1)/2 -
k(k+1)/2$.  If we need to use $k$ spin components to form the ground
state, then for small $\delta H$
every matrix $(H+\delta H + \lambda_i+\delta \lambda_i)$ has
$k$ zero eigenvalues.  So a space of dimension $V(V+1)/2-V$ is
a subspace of a space of dimension $V(V+1)/2-k(k+1)/2$; this is only
possible if $k(k+1)/2 \leq V$.

For non-generic Hamiltonians $H$, we can have $k$ zero eigenvalues of
$H+\lambda_i$, with $k(k+1)/2 > V$.  However, this requires also adjusting
some parameters in the matrix $H$ to produce these extra eigenvalues.
Since we have assumed above that for arbitrary small perturbations of
$H$ there will still be $k$ eigenvalues, we have excluded this non-generic
case.  The case in which $k(k+1)/2>V$ is a case in which the Hamiltonian
has a degenerate ground state, as discussed in the introduction.  However,
the number of spins needed to form the ground state will still
satisfy $k(k+1)/2 \le V$.

Finally, we can check the total number of parameters available to satisfy
the self-consistency equation.  To satisfy the self-consistency equation
we have to satisfy a total of $V$ different equations, one for each
site.  We have $V$ parameters $\lambda_i$ available to satisfy the
equations, so the number of parameters and equations balance.
We have used $k(k+1)/2$ of these parameters to produce the required $k$
zero eigenvalues, so there are $V-k(k+1)/2$ of these parameters left to satisfy
equation (\ref{selcon2}).  Given these $k$ zero eigenvalues, consider
the space of all vectors $\phi_i^{\mu}$, such that
$(H+\lambda_i)_{ij}\phi_j^{\mu}=0$.  Here $i=1...V$ and $\mu=1...k$.
This space is of dimension $k^2$.  
There is a $k(k-1)/2$ dimensional group of rotations
of this space; this leaves a $k(k+1)/2$ dimensional space of distinct
states.  Any state in this space is a
good ground state, so long as it satisfies the constraint on the length
of the spins.  So, we have a total of $V-k(k+1)/2$ parameters
from the different $\lambda_i$ and a total of $k(k+1)/2$ parameters from
the possible $\phi$, so there are $V$ different parameters available to
satisfy the self-consistency equation.  

To give a simple example of this balancing of parameters, 
consider the case $k=1$.  Then, one parameter is required to
produce a single zero eigenvalue.  There are $V-1$ parameters left
over, which are sufficient to make the eigenvector $\phi_i$ have the 
same squared amplitude on each site.  Then, we have one parameter available, 
which is the magnitude of $\phi_i$, to make this amplitude on each site 
equal to unity.

One immediate consequence of this that for sufficiently large
$N$, the system cannot have any metastable states.  Metastable states
are spin configurations which are local minima, but not global minima.
We have shown above that for any $V$, we can construct a ground state using $k$
spin components where $k(k+1)/2\leq V$.  More generally, we have shown
that any local minimum can be constructed using $k$ spin components with
$k(k+1)/2\leq V$.  Let us suppose we have an $N$ component system, where $N$ 
is equal to $2k_{max}$ where $k_{max}$ is the largest $k$ which obeys
$k(k+1)/2\leq V$.  Suppose we have a local minimum of the system, 
$\phi_i^{\mu}$, which is distinct from the ground state $\rho_i^{\mu}$.
Since we have $2k_{max}$ total spin components, we can arrange things so
that $\phi_i^{\mu}$ is nonvanishing only for $\mu=1...k_{max}$, while 
$\rho_i^{\mu}$ is nonvanishing only for $\mu=k_{max}+1...2k_{max}$.
Then, consider the state $\sqrt{1-\delta^2}\phi+\delta\rho$.  As $\delta$
is changed from 0 to 1, this provides a path starting from $\phi$ such
that, near $\delta=0$, the change in energy vanishes to first order
in $\delta$ but is negative  to second order in $\delta$.  This implies that in
fact $\phi_i$ could not have been a local minimum, and that for 
$N\geq 2k_{max}$ there are no local minima other than the ground state.

Given the absence of metastable states shown in this section, and the
uniqueness of the self-consistency equation shown in the last section, it
is reasonable to suppose that replica symmetry breaking is absent in
the system.  Of course, replica symmetry breaking is only defined as
a concept in the replica formalism, but the various concepts of inequivalent
thermodynamic states, and so on, should be absent in the large $N$ problem.
This may be the reason that replica treatments of the large $N$ problem
do not find any instability about the replica symmetric 
solution\cite{oldstuff}.

\begin{figure}[!t]
\begin{center}
\leavevmode
\epsfig{figure=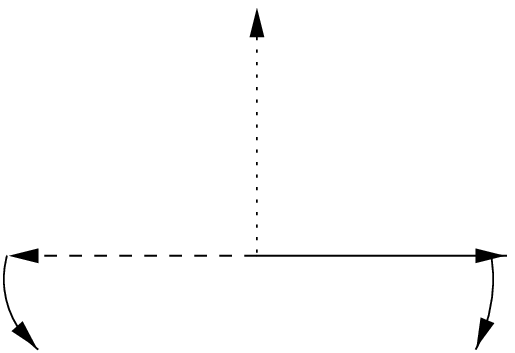,height=4cm,angle=0}
\end{center}
\caption{Illustration of the construction of a series of systems saturating
the bound.}
\label{fig1}
\end{figure}

Finally, we would like to show that the bound derived above for
the maximum $k$ as a function of $V$ is sharp.  That is, for any $k$, we
will construct a system using $V=k(k+1)/2$ spins that employs $k$ spin 
components to form the ground state.  The procedure is discussed below
and illustrated in figure (\ref{fig1}).

First let us look at systems with small $k$.  For $k=1$, we have $k(k+1)/2=1$.
Obviously, a system consisting of one spin always employs one spin component.
For $k=2$, we have $k(k+1)/2=3$.  A system of three spins, all interacting
anti-ferromagnetically with each other, with $H_{ij}=1$ for all $i\neq j$,
will have a ground state using two spin components.  This example was
discussed in the introduction.

To find systems with higher $k$, we will proceed inductively.  Assume that,
for some $k$, there is an $H_{ij}$ acting on a system of
$V_k=k(k+1)/2$ spins such that the ground state uses $k$ spin components.  We
will construct a Hamiltonian acting on a system of $V_{k+1}=(k+1)(k+2)/2$ 
spins which uses $k+1$ spin components.  Note that $V_{k+1}-V_{k}=k+1$.  That 
is, we have an additional $k+1$ spins to use.  

Let the ground state of the system of $V_{k}$ spins be given by
$\rho_i^{\mu}$ where $i=1...V_k$.  Assume, without loss of generality,
that the vectors $\rho_i$ for $i=1...k$ span a $k$-dimensional vector space.
Then, consider the following Hamiltonian
\be
H=\sum\limits_{i,j=1}^{V_k}\phi_i^{\mu} \phi_j^{\mu} H_{ij}
+\sum\limits_{i=1}^{k} \phi_i^{\mu} \phi_{V_k+i}^{\mu}
+ \epsilon \sum\limits_{i=1}^{k} (\phi_i^{\mu}+\phi_{V_k+i}^{\mu})\phi_{V_{k+1}}
\ee
where $\epsilon$ is taken to be sufficiently small.  

The above Hamiltonian can
be thought of as follows: there are still $V_k$ spins with the same interaction
as before.  These spins are those with $i=1...V_k$.  
An additional $k+1$ spins have been added.  First we will consider
the system defined by adding only the first $k$ of these spins, then we
will consider the system defined by adding all $k+1$ of these
spins.

We make each of the first
$k$ of these spins (these are the spins with $i=V_k+1...V_k+k$)
have an anti-ferromagnetic interaction with one of
the spins in the first $V_k$ spins.  
Let us consider this system consisting only of the original $V_k$ spins
and these first $k$ additional spins.  This system has $V_k+k=V_{k+1}-1$ spins.
The ground state of this system is clear.  The first $V_k$ spins are in the 
ground state of the original system ($\phi_i^{\mu}=\rho_i^{\mu}$), 
and the next $k$ spins each point 
opposite to one of the spins in the first $V_k$ spins.  This produces $k$ 
pairs of spins, such that the spins in a pair point in opposite directions.
For example, $\phi_1^{\mu}=-\phi_{V_k+1}^{\mu}$, 
$\phi_2^{\mu}=-\phi_{V_k+2}^{\mu}$, and so on.

Finally, we add the last spin, to produce a total of $V_{k+1}$ spins.
This one additional spin has a weak anti-ferromagnetic interaction with
the $2k$ spins in these $k$ pairs.

Starting with the first $V_k+k$ spins fixed in their ground state, adding
the $V_{k+1}$-th spin does not change the energy at all because the
anti-ferromagnetic interactions cancel.  However, if in a given pair of 
spins, one can slightly bend the pair, so that the two spins form an angle of 
less than $\pi$, then one can gain energy from adding the $V_{k+1}$-th spin.
Since $\epsilon$ is small, the energy gain is small.  However, if the
angle between two spins in a pair is $\pi-\delta$, than the energy gain is
first order in $\delta$, while the energy cost is second order in $\delta$.
So, even for small $\epsilon$, it is advantageous to distort 
the pairs of spins.
As chosen above, the first $k$ spins span a $k$-dimensional space.  

To distort
a pair and gain energy, each spin in that pair has to have some component
of its distortion in a direction opposite to the direction in which the 
$V_{k+1}$-th spin lies.  Also, for small distortions
the direction in which a spin distorts must be 
orthogonal to the direction in which it points, since all spins must have 
fixed length.   So, since each pair must distort in a direction orthogonal
to that in which it lay orignally, and $\phi_{V_{k+1}}$ must 
have some component in the
direction in which that pair distorts, we find that 
the $V_{k+1}$-th spin must have some component which is 
orthogonal to $\phi_i$ for $i=1...k$.  Then, the system of $V_{k+1}$ spins 
must span a $k+1$-dimensional space, as needed.

In figure (\ref{fig1}), we illustrate this procedure, in the simple
case of the transition from $k=1$ to $k=2$, going from $V=1$ to $V=3$.
A solid arrow is used to illustrate the spin in the original system (the
$k=1,V=1$ system), a dashed arrow is used to illustrate the spin which
aligns opposite to that spin, a dotted arrow is used to illustrate the
$V_{k+1}$-th spin (the third spin), and small arrows are used to illustrate
the direction of distortion of the first two spins after the third spin is
added.
\section{Level Repulsion}
An interesting question is the existence of level repulsion for the
large $N$ system.  We know that if $H_{ij}$ is drawn from an ensemble
of matrices that its eigenvalues  will repel each other\cite{rmt}.  This
repulsion is for a noninteracting system.
This can be seen as the result of a Jacobian by writing
\be
\int dH e^{-V \tau {\rm Tr} H^2}=
\int dE \, dO \prod\limits_{i<j} |E_i-E_j| \prod\limits_{i=1}^{V}
e^{-V \tau {\rm Tr} E^2}
\ee
where we have written $H=O^{\rm T} E O$ with $O$ an orthogonal matrix
and $E$ a diagonal matrix of eigenvalues $E_i$ of $H$.  The factor
$|E_i-E_j|$ causes the eigenvalues to repel each other.  This can also
be stated by saying that requiring two eigenvalues of $H$ to be equal to
each other requires tuning two parameters of $H$, instead of one parameter
as might naively be thought.

However, for the interacting large $N$ system we are considering, the
eigenvalues of $H+\lambda_i$ will be different from those of $H$.  In fact,
we know that the solution of the self-consistency equation can force
$k$ of those eigenvalues to scale to zero as $1/\beta$.  This means
that the eigenvalues become very close to each other, differing only
by an amount of order $1/\beta$.  Let us denote the eigenvalues of
$H+\lambda_i$ by $E_{\mu}$, where ${\mu}=1...V$.  Let us have the first
$k$ of these eigenvalues be the ones that scale to zero.
Among those $k$ eigenvalues, there still is a kind of level repulsion.
Consider a problem in the limit $\beta\rightarrow\infty$.  
Then, the solution to the problem
is a configuration of spins $\phi_i^{\mu}$ defining the ground state of the 
system.  These spins span a $k$-dimensional space and so we may assume that
$\mu=1...k$.  By a rotation of the spins, we can further assume that
$\phi_i^{\mu}$ is orthogonal to $\phi_i^{\nu}$ for $\mu \neq \nu$.
That is
\be
\label{orthocon}
\sum\limits_{i=1}^{V} \phi_i^{\mu}\phi_i^{\nu}=0
\ee
Then, by permutation of the spin components $\mu$ we can arrange it that
\be
\label{Edef}
\sum\limits_{i=1}^{V} (\phi_i^{\mu})^2=\frac{N}{\beta E_{\mu}}
\ee
As $\beta\rightarrow\infty$, $E_{\mu}\rightarrow 0$ for
$\mu=1...k$ but $\beta E_{\mu}$ stays non-zero.

Let us show the existence of level repulsion among the $E_{\mu}$ in
a simple model.  Consider a system of three spins, interacting with
some matrix $H_{ij}$.  We know that if $H_{ij}=1$ for all $i\neq j$ then
we have the antiferromagnetic system discussed in the introduction.
It can be checked that the particular ground state of this system given in the
introduction satisfies equation (\ref{orthocon}) and that in this case,
where $\phi_i^{1}=\sqrt{2}(1,-1/2,-1/2)$ and $\phi_i^2=
\sqrt{2}(0,\sqrt{3}/2,-\sqrt{3}/2)$
\be
\sum\limits_{i=1}^{V} \bigl(\phi_i^{1}\bigr)^2=
\sum\limits_{i=1}^{V} \bigl(\phi_i^{2}\bigr)^2
\ee
So then
\be
\beta E_1=\beta E_2
\ee
This system has $k=2$ and both of the eigenvalues which go to zero are equal.
So, we have two levels overlapping.  We would like to know how many
parameters are needed to guarantee that the two levels overlap.  If only
one parameter is needed, then there is no level repulsion.  If two
parameters are needed, then there is level repulsion.  The parameters
we can vary are the different elements of $H_{ij}$; there are three of
these, since $H$ is symmetric and the diagonal elements of $H$ are unimportant.
However, in fact there are only two independent parameters for determining
the ground state of the system, since multiplying all elements of $H$
by a common, positive factor does not change the ground state.

Even if we do not rotate the spin components to make
equation (\ref{orthocon}) true, a simple equation for
$E_1,E_2$ can still be derived.  It can be shown that
\be
\label{tuning}
\frac{1}{E_1}-\frac{1}{E_2}=\sqrt{
\Bigl(\sum_i \bigl(\phi_i^{1}\bigr)^2-\sum_i \bigl(\phi_i^{2}\bigr)^2\Bigr)^2+
\Bigl(\sum_i 2\phi_i^{1} \phi_i^{2}\Bigr)^2}
\ee

Consider matrices $H$ near the uniform antiferromagnetic $H$ considered
above.  There are two independent parameters defining the matrix $H$, if
all we are interested in is ground state properties.  There are also two
independent parameters defining any set of
states equivalent under rotations, since
there are three spins and one overall rotation mode of the three spins.
Let us choose the overall rotation so that we 
continue to pick ground states with $\phi_1=\sqrt{2}(1,0)$.  
Then it is easy to show that small variations in 
parameters in $H$ produces small
changes in $\phi_2,\phi_3$, with no singularities in the Jacobian relating
changes in the two independent parameters of $H$ to the two parameters
defining $\phi_2,\phi_3$.  
So, instead of asking how many parameters of $H$ must be tuned to 
produce $E_1=E_2$, we can ask, when considering level repulsion,
how many of the two independent parameters defining
$\phi_2,\phi_3$ must be tuned to produce $E_1=E_2$.
Looking at equation (\ref{tuning}), we see that we must have
\be
\sum_i \bigl(\phi_i^{1}\bigr)^2-\sum_i \bigl(\phi_i^{2}\bigr)^2=0
\ee
and also
\be
\sum_i 2\phi_i^{1} \phi_i^{2}=0
\ee
This requires tuning both available parameters to satisfy the two requirements.

Then, level repulsion assume an interesting form.  Let us forget
about the underlying Hamiltonian and simply consider arbitrary
spin configurations $\phi_i^{\mu}$.  If we had forgotten about
the orthogonality requirement of equation (\ref{orthocon}) and simply used
equation (\ref{Edef}), it would only require one parameter to make two
levels coincide.  However, either if we rotate the spins to
satisfy equation (\ref{orthocon}) and then use equation (\ref{Edef}),
or if we directly use equation (\ref{tuning}), we find that it requires
additional parameters to make levels coincide, and this implies that there
still is a form of level repulsion in the interacting system.
\section{Spherical Model}
We will discuss a simplification of the large $N$ spin glass.  This is the
spherical model.  This will be a useful simple example.  Here the system
will always condense into one eigenvalue ($k=1$).  The spherical model
is defined by using the Hamiltonian
\be
H=\sum\limits_{i,j}\phi_i \phi_j H_{ij}
\ee
subject to the constraint that
\be
\sum\limits_{i}(\phi_i)^2=V
\ee
Here, there is only one component of $\phi$ on each site, but the
constraint is relaxed to the requirement that the sum of spins over all sites
be equal to $V$.  The problem is again solved by self-consistency, with the
self-consistency equation (again, we rescale $\lambda$ by a factor of $\beta$)
\be
\label{sphsc}
{\rm Tr} (G)=\beta V
\ee
where
\be
G=(H+\lambda)^{-1}
\ee
where $\lambda$ is now independent of the site index $i$.

Consider an ensemble of matrices $H_{ij}$ given by the measure
${\rm d}H \, e^{-V\tau\,{\rm Tr} H^2}$, where $\tau$ is some number of
order unity.  Then it may be shown that, averaged over different $H_{ij}$,
we have\cite{rmt}
\be
\label{gav}
G_{ii}=2\tau \lambda -2\sqrt{\tau^2\lambda^2-\tau}
\ee
Also, we have that, on average,
\be
\label{trgav}
{\rm Tr}(G)=2V\tau \lambda -2V\sqrt{\tau^2\lambda^2-\tau}
\ee
Using this in equation (\ref{sphsc}), we find that
\be
\label{scr}
\lambda=\frac{1}{\beta}+\frac{\beta}{4\tau}
\ee

\begin{figure}[!t]
\begin{center}
\leavevmode
\epsfig{figure=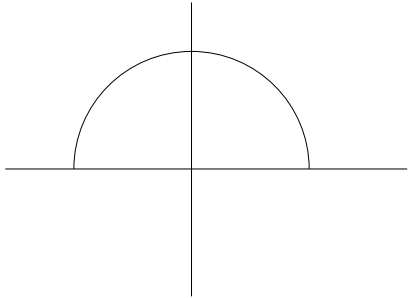,height=4cm,angle=0}
\end{center}
\caption{Wigner semicircle distribution of eigenvalues.  The
horizontal axis is energy, the vertical axis is eigenvalue density.
The semicircle extends from $-\sqrt{\frac{1}{\tau}}$ to 
$\sqrt{\frac{1}{\tau}}$.}
\label{fig2}
\end{figure}

One way of obtaining equation (\ref{trgav}) is by knowing the spectrum
of $H$.  It is known\cite{rmt} that $H$ has a distribution of eigenvalues
obeying the Wigner semicircle, shown in figure (\ref{fig2}).  
The eigenvalues, in the large $V$ limit,
range from $-\sqrt{\frac{1}{\tau}}$ to $\sqrt{\frac{1}{\tau}}$.  The
density of eigenvalues is given by $V\rho(E)$, where 
$\rho(E)=\frac{4}{\pi\tau}\sqrt{\frac{1}{\tau}-E^2}$.
This implies that
\be
\label{trgint}
{\rm Tr}G=\int\limits_{-\sqrt{\frac{1}{\tau}}}^{\sqrt{\frac{1}{\tau}}}
dE\, V \rho(E)\frac{1}{\lambda+E}
\ee
Doing this integral yields equation (\ref{trgav}).

At $\beta=2\sqrt{\tau}$, we have $\lambda=\sqrt{\frac{1}{\tau}}$ from equation 
(\ref{scr}).  At this point the spectrum of
$H+\lambda$ becomes gapless, and a phase transition occurs.  The sum over
eigenvalues needed to define ${\rm Tr} G$ can no longer be approximated
by equation (\ref{trgint}), and now must separately 
include the contribution from the one lowest eigenvalue of $H$.  So,
we have that
\be
{\rm Tr}G=\frac{1}{\lambda+E_{min}}+
\int\limits_{-\sqrt{\frac{1}{\tau}}}^{\sqrt{\frac{1}{\tau}}}
dE\, V \rho(E)\frac{1}{\lambda+E}
\ee
where $E_{min}$ is the lowest eigenvalue of $H$.  The self-consistency
equation (\ref{scr}) can be satisfied in the spin glass phase only if 
$\lambda+E_{min}$ is of order $1/V$.  In this phase, we will have
$\lambda \approx \frac{1}{\tau}$, so that $H+\lambda$ will
be approximately gapless.

More precisely, it is known\cite{bronk,rmt} that the lowest eigenvalue of the
system, $E_{min}$ will lie within a tail that extends to a distance of order
$V^{-1/6}$ below $\frac{1}{\tau}$.  Then, $\lambda=\sqrt{\tau}$ plus a
correction of order $V^{-1/6}$.
\section{Phases of the Large $N$ Infinite Range Model}
The infinite range, large $N$, model of a spin glass exhibits a transition at
a finite $\beta$ from paramagnet to spin glass.  In the paramagnetic
phase, the operator $H+\lambda_i$ has a gap.  In this phase, the
behavior is very similar to that of the spherical model considered above.
In the spin glass phase, 
$H+\lambda_i$ becomes gapless in the thermodynamic (large $V$) limit.
Assuming $k$ components are used to form the ground state of the system,
so that the system condenses into $k$ states, then the operator 
$H+\lambda_i$ will have $k$ eigenvalues of order $k/V$, as well as a 
spectrum
of higher eigenvalues.  First, will briefly discuss the infinite
range model in the paramagnetic phase.

Suppose, as an approximation, that we can look at a system
in which $\lambda_i$ is the same for all values of $i$.  This is
essentially the spherical model considered above.  Let this
value of $\lambda_i$ be denoted $\lambda$.  Then, we find that 
\be
G_{ii}=((H+\lambda)^{-1})_{ii}
\ee
Using equation (\ref{gav}), and requiring that $G_{ii}=\beta$, we find
the same solution of the self-consistency equation as was found in
equation (\ref{scr}).
Further, it may be shown that the mean-square fluctuations in $G_{ii}$ between
different $H_{ij}$, for this constant $\lambda$ are of order $\frac{1}{V}$.

Howver, in the paramagnetic phase, it is not true that $\lambda_i$ is a 
constant.
The slight fluctuations in $G_{ii}$ away from the average
value of equation (\ref{gav}) require $\lambda_i$ to vary from site
to site and sample to sample.  They produce the difference between the 
large $N$ model and the spherical model in this phase.  

We can estimate the fluctuations in $\lambda_i$.  The mean-square 
fluctuations in $G_{ii}$ 
from site to site and sample to sample are of order $1/V$.  So, the 
approximation of taking a constant $\lambda$ will almost satisfy the 
self-consistency equation, but will be slightly in error due to these
fluctuations in $G_{ii}$ away from the average value.  We will
fix this by making a small change in $\lambda_i$.   Let us consider
the change in $G_{ii}$ resulting from a change in $\lambda_j$.  This is equal
to
\be
\label{earlyM}
\delta G_{ii}=G_{ij} \delta\lambda_j G_{ji}
\ee
For $j\neq i$, this vanishes in the thermodynamic limit since 
$G_{ij}G_{ji}$ is of order $1/V$ (this will be discussed in the next section).
For $j=i$, this yields
\be
\label{sfluc}
\delta G_{ii} = \delta\lambda_i \beta^2
\ee
So, if a constant $\lambda$ produces a small error in the self-consistency
equation, we can fix it with a small change in $\lambda_i$ on each
site.  We use equation (\ref{sfluc}) to estimate the change in $\lambda_i$
needed.  Since the mean-square fluctuations in $G_{ii}$ are of order $1/V$,
the mean-square fluctuations in $\lambda_i$ must also be of order $1/V$.
The fact that fluctuations in $\lambda_i$ are small, and that fluctuations
in $G_{ii}$ are small for fixed $\lambda$, is what permits us to approximate
the large $N$ results by the spherical results in this phase.

In the spin glass phase, the spherical model and the
large $N$ model differ greatly.  
In the spherical model, the system condenses only into one eigenvalue.
The eigenvector for this eigenvalue is a randomly chosen vector from a 
$V$-dimensional space of vectors.  So, $G_{ii}$ can be written as the
sum of two parts.  The first is a contribution from the continuum of
eigenvalues above the lowest eigenvalue; this part does not fluctuate
strongly from sample-to-sample
(when the spectrum of $H+\lambda$ becomes gapless, this
part acquires mean-square fluctuations which are much bigger than in the
paramagnetic phase, but which are still small in the large $V$ limit).
The second is a contribution from the lowest eigenvalue.  This
part provides a contribution to $G_{ii}$ of order $V(v_i)^2$, where
$v_i$ is the eigenvector corresponding to this eigenvalue.  This
contribution to $G_{ii}$ has fluctuations from sample-to-sample of order unity,
as $(v_i)^2$ has site-to-site and sample-to-sample fluctuations of order
$1/V$.
So, the spin glass phase and ground state of the large $N$ model
will differ greatly from the spherical model.  The nature
of the ground state is the problem that will be addressed with supersymmetry
techniques later in the paper.  It may be assumed that the state in the
spin glass phase for $\beta \ge 2\sqrt{\tau}$ is very similar to the ground 
state.

Using equation (\ref{energyeq}) and equation (\ref{scr}), we find that
the energy of the system in the paramagnetic phase is given by
\be
-V\frac{\beta}{4\tau}
\ee
At the phase transition point, this becomes $-V\frac{1}{2\sqrt{\tau}}$.
In the spin glass phase, $\lambda \approx \sqrt{\frac{1}{\tau}}$, so
that the energy is given by
\be
V\frac{1}{\beta} - V\sqrt{\frac{1}{\tau}}
\ee

The number of spin components $k$ is in a sense an analogue of
Griffiths effects, known from finite dimensional systems, 
in the infinite range model.  In the rest of this section we will 
pursue this analogy, to help illustrate the importance of the number $k$.  
Consider a finite dimensional system of Bose particles with repulsion.
As the temperature is lowered, there can
be a phase transition to a superfluid phase.  If there is disorder,
there will also be a Griffiths phase, when the system has gapless excitations,
but still has no long-range order.  We can understand the Griffiths phase
as follows: some single particle eigenstate of the finite dimensional
system will be lowest in energy.  The particles will begin to 
condense into this eigenstate.  However, the eigenstate
is localized, and finite in size, so only a microscopic number of particles
can condense into the state before the interparticle repulsion raises
the energy of this state, and particles begin to condense into some
other state.  This leads to a large number of states near zero energy,
as the states lowest in energy have their energy raised by interactions,
and produces the gapless spectrum.
In the infinite range spin glass we consider in this paper, the Griffiths
phase is not a separate phase, but the system still manages to produce a
large number of very low energy states (this number is $k$) by
adjusting $\lambda_i$.

In the spherical model where $\lambda$ is independent of $i$, 
the system always condenses macroscopically
into the lowest eigenstate of the matrix $H$, since 
$H+\lambda$ and $H$ have the same
eigenvectors.  The lowest eigenvalue of $H+\lambda$ will be of order
$1/V$, while the next eigenvalue will be of order $(1/V)^{1/6}$, as
$(1/V)^{1/6}$ is the approximate level spacing in the tail of the
spectrum of $H$, so that the system does not macroscopically occupy
states other than the lowest.

In the large $N$ case, we showed that in the spin glass phase 
the approximation of using constant $\lambda_i$ fails badly.  The system
system must shift $\lambda_i$ to satisfy the self-consistency equation.
In the process of shifting, the system will raise the energy of the lowest
eigenstate of $H+\lambda_i$ more than it raises the energy of the
eigenstates just above it.  This means that several of the eigenvalues
of $H+\lambda_i$ can become of order $1/V$, which leads to $k>1$.  
This process is very similar to the formation of Griffiths states discussed
above.
\section{Correlation Functions}
Let us suppose we have a system with a non-degenerate ground state which
uses $k$ spin components.  We would like to investigate the meaning
of $k$ and how one can see the effects of a given $k$ by looking at the 
correlation functions of the system.  Of course, we have defined $k$
as the dimension of the vector space spanned by the vectors $\phi_i^{\mu}$.
This is some subspace of an $N$-dimensional vector space.  So, if one looks
at all $V$ vectors $\phi_i$, one can determine what $k$ is.  Let us
instead look at correlation functions of the system.

First, let us consider the correlation functions within the paramagnetic
phase discussed in the previous section.  Within the large $N$ formalism,
the correlation function 
\be
\label{corr}
\sum \limits_{\mu}
\langle \phi_i^{\mu} \phi_j^{\mu}\rangle
\ee
is equal to
\be
N G_{ij}
\ee
where the average in equation (\ref{corr}) is a thermodynamic average
for a given Hamiltonian. By self-consistency, we have that
$\sum \limits_{\mu} \langle \phi_i^{\mu} \phi_i^{\mu} \rangle=1$.
Now consider $G_{ij}$ for $i \neq j$, in the paramagnetic phase.
Clearly, this Green's function vanishes after averaging over different
Hamiltonians.  However, we can compute the mean-square fluctuations in
this Green's function, where the fluctuations are from sample-to-sample.

The operator $H+\lambda_i$ has $V$ different eigenvectors.
Denote these eigenvectors by $v^{\mu}_i$, where $\mu=1...V$
now labels the different eigenvectors.  Let each eigenvector have
eigenvalue $E_{\mu}$.  Then
\be
G_{ij}=\sum\limits_{\mu}\frac{1}{E_{\mu}}v^{\mu}_i v^{\mu}_j
\ee
Averaging over different matrices $H_{ij}$ will cause the eigenfunctions
to have random sign on each site, and will cause this quantity to vanish.
Further, 
\be
\label{msq}
G_{ij}G_{ji}=\sum\limits_{\mu,\nu}\frac{1}{E_{\mu}E_{\nu}}
v^{\mu}_iv^{\nu}_i v^{\mu}_jv^{\nu}_j
\ee
Averaging over different matrices $H_{ij}$, we find that the right-hand
side of equation (\ref{msq}) will vanish, except for the terms in which
$\mu=\nu$.  Then we obtain
\be
G_{ij}G_{ji}=\sum\limits_{\mu}\frac{1}{(E_{\mu})^2}
(v^{\mu}_iv^{\mu}_j)^2
\ee
Since we normalize $v^{\mu}_i$ by taking
$\sum\limits_i (v^{\mu}_i)^2=1$, we find that
$(v^{\mu}_i)^2 \approx \frac{1}{V}$.
So, the right-hand side of equation (\ref{msq}) contains $V$ terms of
order $\frac{1}{V^2}$, and so 
$G_{ij}G_{ji}$ is of order $1/V$.

Now, consider the spin glass phase of the system.
There will be $k$ eigenvalues which are of order
$k/V$.  We can again write
\be
\label{g2eq}
G_{ij}G_{ji}=\sum\limits_{\mu}\frac{1}{(E_{\mu})^2}
(\phi^{\mu}_i\phi^{\mu}_j)^2
\ee
as before.  However, we can separate out from this sum the terms
involving the $k$ eigenvalues
$E_{\mu}$ of order $k/V$.  Then, we find that the right-hand side of
equation (\ref{g2eq}) contains $k$ terms of order $\frac{1}{k^2}$.  
So, $G_{ij}G_{ji}$ is of order $\frac{1}{k}$.
\section{Supersymmetric Formalism}
Here we set up a supersymmetric formalism to determine properties of a large 
$N$ system, averaged over disorder.  In the section XI, this formalism will
be used to obtain results on the value of $k$ for the ground state,
the strength of fluctuations in $\lambda_i$, and other results.  
In this section, we
will first derive the formalism in generality for any large $N$ system,
and then discuss simplifications in the low temperature limit.

To average over quenched disorder, many tricks have been invented in
other problems.  The replica trick and Efetov supersymmetry both attempt
to ensure that the partition function of a system, for a fixed realization
of disorder, is equal to unity.  Then, one can average over different
realizations of the disorder directly.  Our goal will be similar.
We wish to find a method of obtaining $\lambda_i$ as a function of
$H$.  We will express this using an integral over all $\lambda_i$, such that
the integral is equal to unity for any $H$.

First we will discuss a naive attempt which does not quite work.  Then
we will introduce a supersymmetry and give the full formalism.

Consider a given $H_{ij}$.  The goal is to find a set of $\lambda_i$
which satisfies the self-consistency equation $G_{ii}=\beta$.
One simple way of implementing this might be to integrate over
all $\lambda_i$ with a set of $\delta$-function
constraints on the Green's function.  
We will also replace equation (\ref{poscon}) with a positivity
constraint on $G$ instead.  Writing these $\delta$-function
constraints as a set of integrals over variables $c_i$, we obtain
\be
\label{wreq}
\int \limits_{G > 0} d\lambda_i \, \frac{dc_i}{2\pi} \,e^{ic_i(G_{ii}-\beta)}
\ee
where $G_{ii}=(H_{ij}+\lambda_i\delta_{ij})^{-1}_{ii}$.  However, the
above equation is not quite right as there is
a non-trivial Jacobian coming from the $\delta$-function.  To find this
Jacobian, assume we have found the set of $\lambda_i$ which satisfy the
self-consistency equation.  Considering a shift $\delta\lambda_i$, we
would find $\delta G_{ii}=G_{ij}\delta\lambda_j G_{ji}$.

Therefore, the integral over $\lambda_i$ and $c$ in equation (\ref{wreq})
is equal to $1/{\rm det}(M_{ij})$, where $M_{ij}$ is the
matrix given by
\be
\label{mdef}
M_{ij}=G_{ij}G_{ji}
\ee
Note that this is not a matrix square, but a square of individual
elements in $G_{ij}$ to produce $M_{ij}$.
To cancel this determinant, we can add an additional set of integrals over
Grassman $\overline a_i,a_i$
variables to equation (\ref{wreq}).  This leads to a new equation
\be
\label{rieq}
\int \limits_{G > 0}
d\lambda_i \, dc_i \, d\overline a_i \, d a_i \,e^{ic_i(G_{ii}-\beta)}
e^{\overline a_i G_{ij}a_j G_{ji}}
\ee

The integral in equation (\ref{rieq}) is equal to unity for any Hamiltonian
$H$.  Then, it is possible to average over Hamiltonians within the equation.
The fact that the integral is equal to unity is due to a supersymmetry.
There are $2V$ bosonic variables and $2V$ fermionic variables.
Similar supersymmetries have been found useful in other systems\cite{parisour}.

It is interesting to compare equation (\ref{rieq}) to equation (\ref{earlyM}).
The equations are similar; as discussed in reference to equation (\ref{earlyM}),
in the paramagnetic phase the matrix $M$ can be treated as
diagonal in the thermodynamic limit.  In the spin glass phase the matrix
$M$ will be very important and will be discussed below and in section XI.

In the infinite range model, we can further simplify equation (\ref{rieq}).
Let us average over Hamiltonians $H$ with Gaussian weight
$e^{-V \tau {\rm Tr} H^2}$.  
We obtain
\be
\label{shift}
\frac{\int \limits_{G > 0} 
d\lambda_i \, dc_i \, d\overline a_i \, d a_i \,
dH e^{-V \tau {\rm Tr} (H^2)}
e^{ic_i(G_{ii}-\beta)}
e^{\overline a_i G_{ij}a_j G_{ji}}}
{\int dH e^{-V \tau {\rm Tr} H^2}}
\ee

For notational simplicity, we will not always write out the factor of
$\frac{1}{\int dH e^{-V \tau {\rm Tr} H^2}}$.  This factor will be
assumed throughout.  At one point in section X, and one point
in section XI, we will need this factor, but elsewhere it will be left out.

For given $\lambda_i$, we can write the integral over $H$ as an integral
over $O$ and $E$ with $H+\lambda_i=O^{\rm T} E O$ (note that the
left-hand side is $H+\lambda_i$, not $H$).  Then $G=O^{\rm T} E^{-1} O$.
Then the integral of equation (\ref{shift}) can be written
\be
\label{orthoshift}
\int \limits_{E \ge 0} 
d\lambda_i \, dc_i \, d\overline a_i \, d a_i 
\, dE \,dO e^{-V \tau {\rm Tr} (H^2)}
\prod \limits_{\mu<\nu} |E_{\mu}-E_{\nu}| \,
e^{ic_i(G_{ii}-\beta)}
e^{\overline a_i G_{ij}a_j G_{ji}}
\ee

For given $k$, we will have $k$ of the eigenvalues $E_{\mu}$ scaling as
$1/\beta$.  In the large $\beta$ limit, these eigenvalues are much smaller
than all other eigenvalues.  It is useful to keep $\beta$ 
finite, but in the large $\beta$ limit we can also assume that $G_{ii}$ is 
made up only of terms due to the $k$ eigenvalues which go to zero, and
ignore contributions to $G_{ii}$ from other eigenvalues.  Then,
equation (\ref{orthoshift}) simplifies.  Let us define $v_i^{\mu}$ as
the eigenvector corresponding to eigenvalue $E_{\mu}$.
In the large $\beta$ limit, $G_{ii}$ does not depend
on $v_i^{\mu}$ for $\mu > k$.  Further, for $\mu=1...k$ and $\nu=k+1...v$,
we have $|E_{\mu}-E_{\nu}|=E_{\nu}$.
Also, we can perform the integral over $\overline a_i,a_i$ to
obtain ${\rm det} (M_{ij})$, where $M_{ij}$ is defined by equation (\ref{mdef}).

For given $k$, we will define a quantity $Z_k$.  This is what is obtained
if equation (\ref{orthoshift}) is evaluated in the sector with $k$ eigenvalues
tending to zero.  Equivalently, 
$Z_k$ is the probability that, if we choose $H$ from
the ensemble at random, we will have $k$ eigenvalues tending to zero. 
It must be the case that $\sum_k Z_k=1$ as a result of the supersymmetry.

So, in the end we have 
\be
\label{final}
Z_k=\int
dv_i^{\mu}dE_{\mu}
\delta(\sum_i (v_i^{\mu})^2-1)
\delta(\sum_i (v_i^{\mu}v_i^{\nu})-1)
Z_k^<[v_i^{\mu},E_{\mu}]
Z_k^>[v_i^{\mu},E_{\mu}]
\ee
where
\be
\label{loreq}
Z_k^<[v_i^{\mu},E_{\mu}]=\int
dc_i \, d\overline a_i \, d a_i 
\, 
\prod \limits_{\mu<\nu} |E_{\mu}-E_{\nu}| \,
e^{ic_i(G_{ii}-\beta)}
\ee
and
\be
\label{upeq}
Z_k^>[v_i^{\mu},E_{\mu}]=\int dH_> \,
d\lambda_i \, 
e^{-V \tau {\rm Tr} (H^2)}
({\rm det}(H_>+\lambda_i)^k
e^{\overline a_i G_{ij}a_j G_{ji}}
\ee
This needs some explanation.  The vectors $v_i^{\mu}$ are the eigenvectors
associated with the eigenvalues $E_{\mu}$ for $\mu=1...k$.  The
eigenvalues $E_{\mu}$
are all of order $1/\beta$.  The Green's function is given by $G_{ii}=
\sum\limits_{\mu=1}^{k} (v_i^{\mu})^2(E_{\mu})^{-1}$.  The matrix $H_>$
is a $V$-by-$V$ matrix, such that
$H_>+\lambda_i$ has only $V-k$ non-zero eigenvalues.  The
zero eigenvectors of $H_>+\lambda_i$ lie in the subspace spanned by the 
$v_i^{\mu}$.
We have $H=H_>+\sum_{\mu}v_i^{\mu}E_{\mu}v_j^{\mu}$.
That is, $H_>+\lambda$ is what results if one looks only at the $V-k$ largest
eigenvalues of $H+\lambda$.  

Equation (\ref{final}) has been written using two equations: equation 
(\ref{loreq}) and equation (\ref{upeq}).
Indeed, equation (\ref{final}) integral almost separates into two parts.  
For convenience, we
will refer to $E_{\mu}$ and $v_i^{\mu}$, for $\mu=1...k$ as the lower
sector, while we will refer to the part of the integral involving
$H_>$ as the upper sector.  Similarly, we may refer to the
upper and lower sectors of the matrix $H+\lambda_i$.  One way
in which the lower and upper sectors communicate is that the eigenvectors
of the lower sector determine the allowed eigenvectors of the upper sector,
which, if $\lambda_i$ is not constant, can change the value of $Z_k^>$.
Also, the determinant due to the integral over Grassman variables depends
on both the upper and lower sector of eigenvalues.

It is interesting to see how the bound $k(k+1)/2 \leq V$ is realized in this
formalism.  We must show that for $k(k+1)/2 \geq V$ the integral of
equation (\ref{final}) vanishes in the large $\beta$ limit.  There
are four pieces to this.  First, the $\delta$-functions, 
$\delta(\sum\limits_{\mu=1}^{k}(v_i^{\mu})^2 (E_{\mu})^{-1} -\beta)$ can be 
rewritten as
$\frac{1}{\beta}\delta(\sum\limits_{\mu=1}^{k}(v_i^{\mu})^2 
(\beta E_{\mu})^{-1} -1)$.
This then contributes a factor of $(\frac{1}{\beta})^V$, as
there are $V$ such $\delta$-functions.  Remember that $E_{\mu}$ scales
as $1/\beta$, so $\beta E_{\mu}$ tends to a constant in the large
$\beta$ limit.

Next, the level repulsion between the first $k$ eigenvalues contributes
a factor of $(\frac{1}{\beta})^{k(k-1)/2}$.  The integral over the first
$k$ values of $E_{\mu}$ contributes a factor of $(\frac{1}{\beta})^{k}$, as
$E_{\mu}$ is of order $\frac{1}{\beta}$.  These two combine to produce
$(\frac{1}{\beta})^{k(k+1)/2}$.

Finally, there is the determinant due to the integral over the Grassman
variables.  This is the determinant of the matrix $M_{ij}$.
First consider the part of this matrix due to the contribution of the
first $k$ eigenvalues to $G$.  This is the determinant of
\be
\label{ginmu}
\sum \limits_{\mu,\nu=1}^{k} v_i^{\mu} v_i^{\nu} (E_{\mu}E_{\nu})^{-1}
v_j^{\mu} v_j^{\nu} 
\ee
This is the sum of $k(k+1)/2$ distinct matrices, each matrix corresponding
to a term in the above sum with given $\mu,\nu$.  It can be seen that
each of these matrices has one nonvanishing eigenvalue which scales as
$\beta^2$ in the large $\beta$ limit.  There is an additional contribution
to the determinant which is equal to
\be
2\sum \limits_{\mu=1}^{k}\sum \limits_{\nu=k+1}^{V} 
v_i^{\mu} v_i^{\nu} (E_{\mu}E_{\nu})^{-1} v_j^{\mu} v_j^{\nu} 
\ee
This is the sum of $k(V-k)$ distinct matrices each with one nonvanishing
eigenvalue of order $\beta$.  There is also a contribution in which
both $\mu$ and $\nu$ are greater than $k$, but this is unimportant.

The determinant of the matrix $M_{ij}$ is the product of the
$V$ eigenvalues of that matrix.  There are at most $k(k+1)/2$ eigenvalues
which are of order $\beta^2$.  These arise from equation (\ref{ginmu}).
The remaining $V-k(k+1)/2$ eigenvalues are of 
order $\beta$.  So the determinant scales as
$\beta^{V+k(k+1)/2}$.

Combining the various factors of $(\frac{1}{\beta})^{V}$,
$(\frac{1}{\beta})^{k(k+1)/2}$, and $\beta^{V+k(k+1)/2}$, we find that
everything cancels.  However, if $k(k+1)/2 > V$, there will be
a problem.  It will not be possible for $M_{ij}$ to have
$k(k+1)/2$ eigenvalues of order $\beta^2$ since $G$ has at most $V$
eigenvalues.  Therefore, if $k(k+1)/2 > V$, equation (\ref{final})
will vanish in the large $\beta$ limit.
\section{A Non-interacting Example}
In the previous section, the integral over $H+\lambda_i$ was written
as an integral over a lower sector with eigenvalues of order $1/\beta$,
and an upper sector.  It will be useful to illustrate this technique
of dividing an integral over random matrices into two sectors, using
a simple example.

Let us consider an integral over all $V$-by-$V$ matrices $H$, with
weight $e^{-V\tau{\rm Tr}H^2}$.  This integral is a Gaussian integral,
and can be performed to yield 
\be
\label{gfac}
\int dH\, e^{-V\tau{\rm Tr}H^2}=
(\frac{\pi}{V\tau})^{-V/2}(\frac{\pi}{2V\tau})^{V(V-1)/2}
\ee
We will separate out $k$ eigenvalues from this matrix, breaking the
integral into two sectors of eigenvalues.  Then we will investigate
the consequences of requiring that the integral over both sectors, including
interaction between the sectors, is equal to equation (\ref{gfac}).

Let us pick out $k$ eigenvalues from $H$, and require them to
be equal to $E_{\mu}$, $\mu=1...k$.  Then, the integral over all matrices $H$ 
can be written as
\be
\label{ovint}
\int dv_i^{\mu}\,dE_1^{\mu} \,dH_> \prod_{\mu}  \Bigl[
{\rm det}(H_>-E_{\mu}) e^{-V\tau E_{\mu}^2} 
\delta(\sum_i (v_i^{\mu})^2-1) \Bigr]
\prod\limits_{\mu<\nu} \Bigl[ |E_{\mu}-E_{\nu}| 
\delta(\sum_i v_i^{\mu}v_j^{\nu}) \Bigr]
e^{-V\tau{\rm Tr}H_>^2}
\ee
where $H_>$ is a $(V-k)$-by-$(V-k)$ matrix.

If $k=1$, the eigenvalue distribution of $H_>$ will be
almost unchanged from a Wigner semi-circle.  Then, the term
$({\rm det}(H_>-E_{\mu}))^k e^{-V\tau E_{\mu}^2}$ is approximately
equal to
\be
(\frac{1}{2\sqrt{\tau}})^{V}e^{-V/2}
\ee
for all $E_{\mu}$ inside the semicircle, $-\frac{1}{\sqrt{\tau}}\leq E_{\mu}
\leq \frac{1}{\sqrt{\tau}}$.
We find
\be
\int dH e^{-V\tau{\rm Tr}H^2} \approx
\int dE 
(\frac{1}{2\sqrt{\tau}})^{V}e^{-V/2}
\int dv_i \delta(\sum_i (v_i)^2-1)
\int dH_> e^{-V\tau{\rm Tr}H_>^2}
\ee
Therefore 
\be
\label{tocancel}
\frac{\int dH e^{-V\tau{\rm Tr}H^2}}
{\int dH_> e^{-V\tau{\rm Tr}H_>^2}}=
\int dE 
(\frac{1}{2\sqrt{\tau}})^{V}e^{-V/2}
\int dv_i \delta(\sum_i (v_i)^2-1)
\ee
The first ratio is equal to
\be
\Bigl(\sqrt{\frac{V\tau}{\pi}} \sqrt{\frac{2V\tau}{\pi}}\Bigr)^{V-1}
\ee
The integral $\int dv_i \delta(\sum_i (v_i)^2-1)$ is approximately
equal to
\be
e^{V/2} (\frac{2\pi}{V})^{V/2}
\ee
It can be verified that, if the last two equations are used in equation
(\ref{tocancel}), that the factors of $e^V$ and $V^V$ cancel, as they
should.  A similar cancellation will be important in the next section.

If $k>1$, the term $\prod\limits_{\mu<\nu} |E_{\mu}-E_{\nu}|$ might
seem to make equation (\ref{ovint}) greatest when the eigenvalues
$E_{\mu}$ are well separated.  However, we know that the original matrix
$H$ has $V$ eigenvalues, with level density of order $V$ and separation
between levels of order $1/V$.  For well separated levels, with
$|E_{\mu}-E_{\nu}|>>1/V$, the effect of
level repulsion is negligible.  The solution to this problem lies in the
determinants $\prod_{\mu} {\rm det}(H_>-E_{\mu})$.  For a given value
of $E_{\mu}$ the determinant, ${\rm det}(H_>-E_{\mu})$,
will slightly alter the level distribution of $H_>$.
This will produce a slight decrease 
in level density around $E_{\mu}$.  This will then, as a result of the
determinant,
${\rm det}(H_>-E_{\nu})$, produce an effective attraction between levels
that cancels out $\prod\limits_{\mu<\nu} |E_{\mu}-E_{\nu}|$ if
$|E_{\mu}-E_{\nu}|$ is much greater than $1/V$.

Although this discussion seems elementary, it is worth reviewing this
for what will happen in the next section.  Then, the lower sector of
$H+\lambda_i$ will have $k$ eigenvalues all within order $1/\beta$ of each
other.  This will produce a change in the level density of $H_>$, and
will change the determinant $({\rm det}(H_>+\lambda_i))^k$ away from the naive
value one would obtain by assuming that $H_>$ has a Wigner semicircle
density of eigenvalues.  The true level density of $H_>$ will be evaluated
in the next section and in the appendices.  Since $k$ will be much greater
than one, this change in level density will be very important.
\section{Ground State for the Large $N$ Infinite Range Model}
Having set up the supersymmetric formalism in the previous section,
we will use this formalism to determine the most likely value of $k$ and
the strength of fluctuations in $\lambda_i$.  It will be shown that the
system will obey $k\propto V^{2/5}$.  Further, it will
be shown that the mean square fluctuations in $\lambda_i$ are equal to
$\frac{1}{k\tau}$.  We will also derive a slight change in the
average value of $\lambda$ from $\sqrt{\frac{1}{\tau}}$, and a
slight gap in the low energy excitations.

The starting point for the calculation is equation (\ref{final}).  We
will evaluate this integral for different $k$ and use this to determine
the most likely $k$.  For any given $k$, we can write the value of this
integral as $Z_k=e^{S_k}$.  Below we will evaluate various contributions
to $S_k$.  Due to the complexity of the calculations, we will only evaluate
terms in $S_k$ to order 
$kV, V{\rm log} V, V {\rm log} k, V, k^2 {\rm log} V, $ and $k^2 {\rm log} k$.
There will also be terms in $S_k$ of order $k^2,V/\sqrt{k}$, and smaller,
which we will not evaluate.
All the terms of order $kV, V {\rm log k}, V {\rm log V},$ and $V$ will cancel.
It will turn out that the other terms
do not cancel; given these terms, we will show at the end of the section
that $k$ is of order $V^{2/5}$, with the exact ratio $k/V^{2/5}$ undetermined
without a computation of terms of order $k^2$ and $V/\sqrt{k}$ and in $S_k$.
Note that for $k$ of order $V^{2/5}$, then $k^2$ and  $V/\sqrt{k}$ are
of the same order.

The procedure to evaluate equation (\ref{final}) will be to start
with all $\lambda_i$ equal to each other.  We will then write 
$\lambda_i=\lambda$.  In this case, we will evaluate the integral over the 
various eigenvalues $E$ and eigenvectors $v$ to obtain the result.  Then we
will consider the effect of site-to-site fluctuations in $\lambda_i$ at the 
end.  From this we will get an effective action for $\lambda$ which will tell us
the average value of $\lambda_i$ and the strength of fluctuations about this
average.

The rest of this section will be divided into four subsections.
First there will be a calculation of the contribution to $Z_k$ from
the upper sector.  Then there will be a calculation of the contribution
to $Z_k$ from the lower sector.  These two calculations will be for the
case with
all $\lambda_i$ equal to a constant $\lambda$.  In the third subsection, we
will consider the effect of permitting $\lambda_i$ to vary from site to site.
In the final subsection, we will put the calculations of the first three
subsections together and obtain final results.
\subsection{Upper Sector}
For constant $\lambda_i=\lambda$, $Z_k$ is completely independent of the
eigenvectors $v_i^{\mu}$ for $\mu=k+1...V$.  $Z_k$ does still depend
on $E_{\mu}$ for $\mu=k+1...V$.  The matrix $H+\lambda$ has
$k$ eigenvectors near zero and $V-k$ other eigenvectors which are all
positive.  The average value of $\lambda$ is determined by a balance
of two effects.  First, there is a term in the
probability distribution for $\lambda$ (when all $\lambda_i$ are set equal to 
$\lambda$) like $e^{-k V {\tau} \lambda^2}$.  This term simply arises from 
the requirement that $H$ must 
have $k$ eigenvectors equal to $-\lambda$ and 
from the term in equation (\ref{upeq}), $e^{-V \tau{\rm Tr }E^2}$,
where $E$ are the eigenvalues of $H$.
This term favors a small value of $\lambda$.
There is also a term due to the integral over $E_{\mu}$ for $\mu=k+1...V$.
The requirement that these eigenvalues all be positive, and the effect
of level repulsion between these eigenvalues and the first $k$ eigenvalues,
tend to favor a larger value of $\lambda$.

For example, in the simple non-interacting problem of the previous
section, when we removed one eigenvalue, the level repulsion
and the Gaussian weight canceled, and the eigenvalue could have been
found anywhere between $-\sqrt{\frac{1}{\tau}}$ and $\sqrt{\frac{1}{\tau}}$.
In the case considered in this section, the effects of level repulsion
will be stronger, as $k$ eigenvalues will coalesce.  Further, the positivity
constraint must be satisfied.  We will also find from this a slight
difference between $\lambda$ for the large $N$ model and $\lambda$ for
the spherical model, which will imply a shift in energy from equation
(\ref{energyeq}).

The integral over the upper sector of eigenvalues is performed in
appendix A.  There, it is shown that the eigenvalue distribution of the
upper sector is slightly distorted from the Wigner semicircle, and that
$\lambda$ is slightly shifted from $\frac{1}{\sqrt{\tau}}$ to
$\frac{1}{\sqrt{\tau}}-\frac{3}{4}\frac{1}{\sqrt{\tau}}2^{1/3}
(\frac{k}{V})^{2/3}$.  If the eigenvalue
distribution of the upper sector were unchanged from a Wigner semicircle
and $\lambda=\sqrt{\frac{1}{\tau}}$, the
determinant $({\rm det}(H_>+\lambda))^k$ would equal
\be
\label{naicon}
(\frac{1}{2\sqrt{\tau}})^{k(V-k)}e^{k(V-k)/2}
\ee
Also, $e^{-kV\tau\lambda^2}$ would equal $e^{-kV}$.
However, the slight change in the eigenvalue spectrum, and slight shift in
$\lambda$ lead to
\be
\label{con1}
({\rm det}(H_>+\lambda))^k e^{-kV\tau\lambda^2}=
(\frac{1}{2\sqrt{\tau}})^{k(V-k)} e^{k(V-k)/2} e^{-kV}
(\frac{k}{V})^{-\frac{2}{3} \frac{k^2}{2}}
\ee
as can be shown using the results in appendix A and calculating the
determinant by integrating the eigenvalue density from appendix A.

\begin{figure}[!t]
\begin{center}
\leavevmode
\epsfig{figure=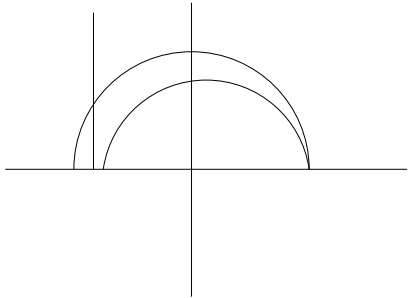,height=4cm,angle=0}
\end{center}
\caption{Altered level distribution of $H+\lambda$ in 
large $N$ system.  The spike
contains $k$ eigenvalues, the smaller semicircle contains $V-k$ eigenvalues.
Superimposed is a semicircle of a matrix containing $V$ eigenvalues.
The horizontal axis is energy, although the scale is shifted so that the
spike is at zero energy.}
\label{fig3}
\end{figure}

In appendix A we calculate $\lambda$ and $a$, where $a$ is 
the lowest eigenvalue of $H_>$.  The sum, $\lambda+a$
gives a gap to excitations.  The gap is equal to
$\frac{1}{4}2^{1/3}\frac{1}{\sqrt{\tau}}(\frac{k}{V})^{2/3}$ and will vanish 
as a power law in $V$.
In figure (\ref{fig3}) we illustrate the changed eigenvalue distribution,
superimposing the Wigner semicircle of the original matrix $H$ for comparison.
The gaps in the figure are exaggerated and are not to scale; the figure
is approximate only.

The factor of $\frac{2}{3}$ in the various above equations should not
be a surprise.  For a matrix with a Wigner semicircle distribution
of eigenvalues, there is a square-root singularity in the eigenvalue
density near the end of the semicircle.  This means that the lowest
$k$ eigenvalues range in energy up to a distance of order $(k/V)^{2/3}$ from
the end of the semicircle.  Having changed the distribution of eigenvalues
from the Wigner semicircle
so that the lowest $k$ eigenvalues all lie at zero energy, the eigenvalue
density of the rest of the matrix will adjust,
but once one looks at energies several times $(k/V)^{2/3}$ from the
tail, the change in the eigenvalue distribution is small as
a result of screening.  This argument gives the scaling of the gap.

To understand the difference between equation (\ref{naicon}) and
equation (\ref{con1}), 
we must calculate $e^{-kV\tau\lambda^2} ({\rm det}(H_>+\lambda)^k$,
for given $\lambda$.
The naive evaluation of this is what gave
equation (\ref{naicon}).  The evaluation of the determinant is very
much like evaluating the potential energy of a system of charges
interacting with a logarithmic Coulomb interaction.  We can then write
the true eigenvalue distribution of $H_>$ as the sum of two pieces:
a Wigner semicircle extending from $-\sqrt{\frac{1}{\tau}}$ to
$\sqrt{\frac{1}{\tau}}$, and a negative term extending to a distance
of order $(k/V)^{2/3}$, with the integral of the negative term being $k$.
The negative term can be thought of as a ``hole".  

The interaction with the Wigner semicircle, multiplied by 
$e^{kV\tau\lambda^2}$, is independent of $\lambda$, for $\lambda$ inside
the semicircle, and the product is equal to equation (\ref{naicon}).
So, we can think of equation (\ref{con1}) as arising from equation 
(\ref{naicon}), multiplied by a contribution arising from the interaction
of the $k$ levels of $H$ at $-\lambda$ with the ``hole".  This gives
equation (\ref{con1}).

Finally, since we are considering the integral over $H_>$, recall that in 
equation (\ref{final}), we have normalized the integral
by
\be
\label{ratio1}
\frac{1}{\int dH \,e^{-V\tau{\rm Tr} (H^2)}}
\ee
In the calculation of equation (\ref{con1}) and of
appendix A, when we integrate over $H_>$, we gain an extra factor of
\be
\label{ratio2}
\int dH_> \,e^{-V\tau{\rm Tr} (H^2)}
\ee
The ratio of these produces a factor of
\be
\label{ratcon}
\sqrt{\frac{V\tau}{\pi}}^{k} \sqrt{\frac{2V\tau}{\pi}}^{kV-k^2/2-k/2}
\ee
This is similar to the calculation in section X.
\subsection{Lower Sector}
The rest of equation (\ref{final}) involves the lower sector of eigenvalues.
Let us write 
\be
\phi_i^{\mu}=v_i^{\mu}\sqrt{\frac{1}{E_{\mu}}}
\ee
Here, $\phi_i^{\mu}$ is a ground state of the system, up to a change
in normalization (compare to equation (\ref{normali}).
Further, let us change variables and write $E_{\mu}=\rho_{\mu}^{-1}$.
Then, the lower sector of the integral can be written as
\be
\label{lowsec}
\int d\rho_{\mu} \, d\phi_i^{\mu} \, 
\prod\limits_{\mu}\rho_{\mu}^{-V/2-1-(k-1)/2}
\delta\Bigl(\sum\limits_{\mu} (\phi_i^{\mu})^2-\beta\Bigr)
\delta\Bigl(\sum\limits_i (\phi_i^{\mu})^2-\rho_{\mu}\Bigr)
\prod\limits_{\mu<\nu}
\delta\Bigl(\sum\limits_i (\phi_i^{\mu}\phi_i^{\nu})\Bigr)
|\rho_{\mu}-\rho_{\nu}|
{\rm det}(M_{ij})
\ee

If we ignore the determinant in equation (\ref{lowsec}), the
integral over $\phi$, at fixed $\rho$, can be performed using a series
of Lagrange multipliers to implement the $\delta$ functions.
We can take the integral
\be
\int d\phi_i^{\mu}
\prod\limits_{\mu}
\delta(\sum\limits_{\mu} (\phi_i^{\mu})^2-\beta)
\delta(\sum\limits_i (\phi_i^{\mu})^2-\rho_{\mu}))
\prod\limits_{\mu<\nu}
(\delta(\sum\limits_i (\phi_i^{\mu}\phi_i^{\nu}))
\ee
and write it as
\be
\label{csadq}
\int d\phi_i^{\mu}\,
\frac{dc_{\mu\nu}}{2\pi}\, \frac{dc_{\mu}}{2\pi}\, \frac{dc_i}{2\pi}\,
e^{\phi_i^{\mu} (
ic_i\delta_{\mu\nu}+
ic_{\mu\nu}+ic_{\mu}
)\phi_i^{\nu}}
e^{-i\rho_{\mu}c_{\mu}-i\beta c_i}
\ee

We can integrate over the fields $\phi_i^{\mu}$, to obtain an action
for the Lagrange multiplier fields $c$.  The integral over $c$
can then be handled by the saddle point method.  One possible saddle point is
$c_{\mu\nu}=0, c_{\mu}=iE_{\mu}V/2, c_i=0$.  There is a zero mode
for fluctuations about this saddle point which enforces the constraint
$\sum_{\mu} \rho_{\mu}=\beta V$.  This zero mode is a mode in which
all the $c_i$ are increased by some constant, while all the $c_{\mu}$ are
decreased by the same constant so that $c_{\mu}+c_i$ is unchanged for all
$\mu$ and $i$.

At the saddle point, we find that the integral given by equation (\ref{csadq})
is equal to
\be
\label{con2}
e^{Vk/2}
\prod\limits_{\mu} \Bigl(\frac{2\rho_{\mu}\pi}{V}\Bigr)^{V/2}
\ee

For $V$ much greater than $k$, we can look at other fluctuations about
the saddle point in a Gaussian approximation.  The fluctuations in
$c_{\mu\nu}$ contribute a factor of
\be
\label{con3}
\int
\frac{dc_{\mu\nu}}{2\pi}\,e^{-\frac{1}{2V}c_{\mu\nu}^2\rho_{\mu}\rho_{\nu}}=
(\frac{2\pi}{V})^{+k(k-1)/4}
(2\pi)^{-k(k-1)/2}
\prod\limits_{\mu}(\rho_{\mu}/V)^{-(k-1)/2}
\ee
The fluctuations in $c_i$ contribute a factor of 
\be
\label{con4}
\int
\frac{dc_i}{2\pi} \, e^{-\frac{1}{4}c_i^2\sum_{\mu}(2\rho_{\mu}/V)^2} =
(\frac{4\pi}{\sum_{\mu}(2\rho_{\mu}/V)^2})^{V/2}(2\pi)^{-V}
\ee
The fluctuations in $c_{\mu}$ can be ignored, as there are only $k$ such
terms fluctuating, and this would only lead to corrections to $S_k$ of
order $k$.  We are not considering terms in $S_k$ that small.

We will now evaluate the contribution to $Z_k$ from
${\rm det}(M_{ij})$.  This determinant involves both the lower and
upper sector of eigenvalues.  As discussed
in section IX, this matrix has $k(k+1)/2$ eigenvalues of order $\beta^2$
and $(V-k(k+1)/2)$ eigenvalues of order $\beta$.  The trace of the
matrix is $V\beta^2$.  So, the $k(k+1)/2$ largest eigenvalues can be
at most of order $\frac{V\beta^2}{k(k+1)/2}$.  Let us assume that in fact 
each eigenvalue is approximately $\frac{V\beta^2}{k^2}$; making
the approximation only leads to errors of order $k^2$ in $S_k$.

Since we will find in the end that $k \propto V^{2/5}$, we have
$V-k(k+1)/2 >> k(k+1)/2$.  So, to find the product of the other
$V-k(k+1)/2$ eigenvalues, we can simply throw out terms in $M_{ij}$ of
order $\beta^2$ (these are terms shown in equation (\ref{ginmu})) and calculate 
the determinant of the remaining $V$-by-$V$ matrix.  We find
\be
G_{ij}G_{ji}=2\sum\limits_{\mu=1}^{k}\sum\limits_{a=k+1}^{V}
\frac{1}{E_{\mu}E_{a}}
v^{\mu}_iv^{\nu}_j v^{a}_iv^{a}_j=
2\sum\limits_{\mu=1}^{k}\sum\limits_{a=k+1}^{V}
\frac{1}{E_{a}}
\phi^{\mu}_i\phi^{\mu}_j v^{a}_iv^{a}_j
\ee
Here, $E_{\mu}$ is of order $1/\beta$, while $E_{a}$ is from the upper
sector of eigenvalues.  Averaging over $\phi_i^{\mu}$, we can ignore
terms in which $i \neq j$ and obtain
\be
\label{quieq}
G_{ii}G_{ii}\approx 2\sum\limits_{a=k+1}^{V}
\beta\frac{1}{E_{a}} v^{a}_iv^{a}_j
\ee
This is a Green's function of the matrix $H_>$.  The Green's $G_{ii}$ 
function of $H_>$ is approximately  $2\sqrt{\tau}$ and so
equation (\ref{quieq}) is equal to
\be
4\beta\sqrt{\tau}
\ee
Putting these together, we obtain
\be
\label{con5}
{\rm det}(M_{ij})\approx
(\frac{V\beta^2}{k(k+1)/2})^{k(k+1)/2}(\beta 4\sqrt{\tau})^{V-k(k+1)/2}
\ee
There will be corrections to this which lead to corrections to
$S_k$ of order $V/\sqrt{k}$.  These corrections will be discussed at
the end of this section and in appendix B.

We now need to know what the spectrum of $E_{\mu}$ is, for $\mu=1...k$,
to determine what some of the contributions calculated above are equal to.
The exact spectrum is difficult to determine, but we can determine
enough for our purposes.
There is an effect of level repulsion which tends to
push the energies apart, but it may be seen that the other terms
in equation (\ref{lowsec}) are greatest when the energies are all
equal to $k/(V\beta)$.  The effects of
level repulsion will push the $E_{\mu}$ apart, but on the other
hand if we considered only
the effects of level repulsion, ignoring all other effects except
the requirement that all eigenvalues be positive, we still
want all $E_{\mu}$ to be of order $k/(V\beta)$, up to factors of
order unity.  So, considering all effects, we must have that
$E_{\mu}$ is of order $k/(V\beta)$.

Also, we would like to know how large $|E_{\mu}-E_{\nu}|$ is.  
Although each $E_{\mu}$ is of order $k/(V\beta)$, the separation between
energy levels may be much smaller, as the contribution
of equation (\ref{con4}) depends on $E_{\mu}$ and is greatest when all
$E_{\mu}$ are equal.  There are other terms which have a similar effect
and are also greatest when all $E_{\mu}$ are equal.  However, we will
now show that despite this $|E_{\mu}-E_{\nu}|$ is still of order
$k/(V\beta)$.
It can be shown that, if equation (\ref{con4}) is expanded out in 
$\delta E_{\mu}=E_{\mu}-k/(V\beta)$, that there will be term of the form
\be
\label{abveq}
e^{-\frac{1}{4}V(\frac{\delta_{\mu}}{\beta V})^2}
\ee
Thus, the energy levels $E_{\mu}$ repel each other due to the absolute
value, $|E_{\mu}-E_{\nu}|$, but are confined by a potential that, for
$E_{\mu}$ approximately equal to $k/(V\beta)$, is given by equation 
(\ref{abveq}).  If we have $k$ levels, $E_{\mu}$, repelling 
each other, in a confining potential
given by equation (\ref{abveq}), then for $k$ of order $V^{1/3}$,
or greater, we find that
the separation between levels is of order $k/V\beta$.  So, we can
assume that $|E_{\mu}-E_{\nu}|$ is of order $k/(V\beta)$ since we
will find in the end that $k=V^{2/5}>V^{1/3}$.

Putting all this together, we obtain a contribution to equation 
(\ref{final}) from the lower sector given by
equation (\ref{lowsec}) and a contribution from the upper sector
given by equations (\ref{con1},\ref{ratcon}), as well as the determinant
of equation (\ref{con5}).  All this is at fixed $\lambda$.
Equation (\ref{lowsec})
includes a number of terms due to the change of variables and from
the level repulsion, as well as contributions evaluated in equations
(\ref{con2},\ref{con3},\ref{con4}).  If we take typical
$\rho_{\mu}$ to be of order $V\beta/k$ as discussed in the above
paragraphs on level repulsion, we find
\be
\label{zk1}
Z_k\approx(\frac{V}{k^2})^{k^2/2} (\frac{k}{V})^{+\frac{1}{3}\frac{k^2}{2}}
(\frac{4k\tau}{\pi})^{V/2}
\ee
\subsection{Site to Site Fluctuations in $\lambda_i$}
The above calculation was performed for given $\lambda$.  Finally, we
must obtain the strength of fluctuations in $\lambda$, and perform the
integration over $\lambda$.  We will now obtain an effective
action for $\lambda$.  Let $\lambda_i=\lambda+\delta\lambda_i$,
where $\lambda$ is the average value of $\lambda_i$ calculated above.
Then, $\sum_i \delta\lambda_i=0$; this is important and will be discussed
more below.
The action for $\lambda_i$ arises from the change in 
$({\rm det}(H_>+\lambda))^k$,
as well as the change in $H$ required to ensure that $H+\lambda$ still
has $k$ zero eigenvalues.

First, let us evaluate the change resulting from the determinant.
We have 
\be
({\rm det}(H_>+\lambda))^k=
e^{k {\rm Tr \, log}(H_>+\lambda_i)}
\ee
Expanding the log in the above equation in $\delta\lambda_i$ we obtain 
to second order
\be
\label{lamdac}
e^{k {\rm Tr \, log}(H_>+\lambda)}
e^{-k \frac{1}{2}  
G_{ii}^2\delta\lambda_i^2}=
e^{k {\rm Tr \, log}(H_>+\lambda)}
e^{-2 k \tau \delta\lambda_i^2}
\ee

Finally, remember that $H+\lambda_i$ must have $k$ zero eigenvalues.
For $\lambda_i=\lambda$, a constant, this contributed a term
$e^{-kV\tau\lambda^2}$ as found in equation (\ref{con1}).
Let us write $H$ in a basis of the eigenvectors of $H+\lambda_i$.
Then, since $(H+\lambda_i)v^{\mu}=0$, for $\mu=1...k$, we must have
$H_{\mu,a}=\sum_i v_i^{\mu} \lambda_i v_i^{a}$, for $\mu=1...k$ and
$a=1...V$.  Also $H_{a,\mu}=\sum_i v_i^{a} \lambda_i v_i^{\mu}$.
Then, we must evaluate
\be
e^{-V\tau\sum\limits_{\mu=1}^k\sum\limits_{a=1}^V(H_{\mu,a}^2+H_{a,\mu}^2)}=
e^{-2V\tau\sum\limits_{\mu=1}^{k}\sum\limits_{a=1}^{V}
(\sum_i v_i^{a} \lambda_i v_i^{\mu})^2}
\ee
Since $(v_i^{\mu})^2$ and $(v_i^{a})^2$ are approximately $1/V$, this is
\be
\label{lamac}
e^{-2k\tau\delta\lambda_i^2}
\ee

Multiplying equations (\ref{lamdac},\ref{lamac}), and integrating over
$\lambda_i$, we get find an extra contribution to equation (\ref{final}) 
\be
\label{excon}
(\frac{\pi}{4k\tau})^{V/2}
\ee
Corrections to this will lead to corrections to $S_k$
of order $V/\sqrt{k}$.  These will be discussed later.

The result from equations (\ref{lamdac},\ref{lamac}) is interesting in
that it requires that $\sum_i \delta\lambda_i=0$.  
If we evaluated the shift
in determinant for constant shift in $\lambda$ (all $\delta\lambda_i$ equal
to the same constant), it would exactly cancel
the change in $e^{-V\tau H^2}$.  However, here we would find that the
perturbative method used to obtain equation (\ref{lamdac}) was not valid,
and that constant shift in $\lambda$ could violate equation (\ref{poscon}).
Further, we would not have the factor of two in equation (\ref{lamac})
for constant shift in $\lambda$.  The factor of two arises since the shift 
due to $\delta\lambda_i$ involves off-diagonal terms in $H$, while
a constant shift in $\lambda$ would involve diagonal terms in $H$.
Effectively, if we look at fluctuations in $\lambda_i$, the mode
in which all $\lambda_i$ fluctuate together is very different
from the other modes in which $\sum_i\delta\lambda_i=0$.  The mode in 
which all $\lambda_i$ fluctuate together will be discussed in the next
section; this mode is important for considering fluctuations in total energy.
\subsection{Results}
The final result, combining equations (\ref{zk1}) and (\ref{excon}), is 
that $Z_k$ is of order
\be
\label{zkfrk2}
(\frac{V}{k^2})^{k^2/2} (\frac{k}{V})^{+\frac{1}{3}\frac{k^2}{2}}
\ee
In obtaining this result a large number of terms have canceled.  It may be
verified that all terms with an exponent of order $kV$ or $V$ cancel.  Further,
it may be verified that $\tau$ cancels exactly, as required.

Equation (\ref{zkfrk2}) is maximized for 
\be
\label{propeq}
k \propto V^{2/5}
\ee
There will also be terms contributing
to $Z_k$ which look like $e^{a k^2}$, for some constant $a$, as
well as exponential of lower powers of $k^2$ and powers of $V$.  To
evaluate these terms is beyond the scope of the present calculation.
These terms
can shift the constant of proportionality in equation (\ref{propeq}), but
cannot change the power law dependence of $k$ on $V$.  It would be nice
if a more detailed calculation could obtain the exact proportionality
constant, as well as checking that $\sum_{k} Z_k=1$.  The
requirement that $\sum_{k} Z_k=1$ can be verified here only so far as noting
that for $k \propto V^{2/5}$ that $Z_k$ does not have any terms in it
like $e^{k^2 {\rm log} k}$ or like $e^V$, 
but we cannot check for cancellation of contributions to $S_k$ of order 
$k^2=V^{4/5}$ or lower order.

Further, there will be contributions to $S_k$ of order $V/\sqrt{k}$ that would
arise from a more careful calculation of ${\rm det}(M_{ij})$ and from
a better calculation of the integral over fluctuations in $\lambda_i$.  We
will discuss these terms in appendix B, and show that they are in
fact of order $V/\sqrt{k}$.  For $k=V^{2/5}$, these terms will be of
the same order as the $k^2$ contributions to $S_k$.  For $k<V^{2/5}$, these
terms will be the most important contribution to $S_k$, and since they
are negative, we will find that $S_k$ is negative and $Z_k$ is
much less than 1; for $k>v^{2/5}$, 
equation (\ref{zkfrk2}) will give the most important contribution and
will also lead to $Z_k$ being small.  Therefore, although terms
of order $V/\sqrt{k}$ can change the proportionality constant in 
equation (\ref{propeq}), they
cannot change the power law dependence of $k$ on $V$.  These terms can
also lead to a change in the gap, $\lambda+a$,
derived in appendix A, but will not change the scaling of the gap with $V$.

The results on the ground state can be summarized by equation
(\ref{propeq}) for the value of $k$, equation (\ref{lamac}) for the
fluctuations in $\lambda_i$, and the calculation in appendix A for
the average value of $\lambda$, and the weak gap to excitations above the
ground state.
\section{Discussion of Results on the Ground State and Connection to
Properties in the Spin Glass Phase}
The two most important results are the value of $k$ and the strength of
fluctuations in $\lambda$.  We will discuss these results in this section
and try to interpret them.  Next we will discuss fluctuations in
the total energy of the system, as discussed at the end of section XI, 
subsection C.  The shift in $\lambda$ is also
interesting all will be discussed next.  Finally, we will argue that
the results of the previous section are also applicable to the
spin glass phase as well as the ground state.

The value of $k$ is perhaps somewhat
suprising.  One might have expect $k \propto \sqrt{V}$ as we have the bound 
that $k(k+1)/2\leq V$.  However, imagine considering a different problem
from that of finding ground states of the system.  Suppose we had instead
looked for stationary states.  These are configurations of
the spins such that the energy does not change for small changes in
spin configuration.  The problem of finding stationary
states is the same as the problem of finding ground states, except
that equation (\ref{poscon}) does not need to be satisfied any more.
Then, consider a solution in which all $\lambda_i$ are equal to some
constant $\lambda$.  Now, we can take $\lambda$ to lie in the middle of
the Wigner semicircle, instead of near the edge.  Then, the solution
of the problem would proceed exactly as before, except that the corrections
to the determinant calculated in appendix A would be of order
\be
(\frac{V}{k})^{k^2/2}
\ee
without the factor of 2/3 in the exponent (compare to the discussion of
the origin of the factor 2/3, as this factor arises near the edge
of the semicircle).  In this case, it may be verified that
$Z_k$ is greatest for $k^2$ of order $V$.  

Since the same bound on $k$ holds for stationary states as for ground
states, the most likely value of $k$ for stationary states is not
surprising, as here $k^2$ is of order $V$.  
However, one can perhaps argue that typical $k$ should
be less for a ground state than for a stationary state.  From most
stationary states one can reduce the energy by aligning some spins and
reducing $k$, bringing one closer to the ground state.
This argument is of course very crude, but it is an attempt to
interpret the result for $k$, especially since the calculation of
section XI and appendix A shows that the result for $k$ depends crucially
on the square-root singularity of the density of states of $H$.

In fact, it is not surprising that the square-root singularity should
lead to a reduction in $k$.  We expect that $\lambda$ will be near
$\sqrt{\frac{1}{\tau}}$.  The system must take $k$ eigenvalues
near $-\sqrt{\frac{1}{\tau}}$ and bring them to zero energy, by tuning
$\lambda_i$.  The square-root singularity implies that there is a
lower density of eigenvalues near the end of the semi-circle, and so it
is more difficult to bring $k$ eigenvalues to zero energy.  When looking
for stationary states, instead of ground states, we can take $\lambda$
near the middle of the semicircle, and there are many more eigenvalues
around to bring to zero energy, so $k$ can be bigger.

More interesting is the strength of fluctuations in $\lambda$.  We have
shown that the mean square fluctuations are of order $1/k$.  The fluctuations
in $\lambda$ on a given site measure the fluctuations in the energy of the
bonds connecting that site to other sites.  Each site is connected to
$V$ other sites.  The term in the Hamiltonian connecting the two sites has
mean square of order $1/V$.  However, the energy due to a bond connecting two 
sites is determined both by the strength of the bond and the correlation
between the two spins on the sites.  Above, we estimated that the mean
square correlation function between two sites is of order $1/k$.
Putting all this together, the energy on a given site is the sum of
$V$ terms, each with mean square fluctuations of order $\frac{1}{Vk}$.
So, it is not surprising that the mean square fluctuation in $\lambda$
for a given site would be of order $1/k$.

The total ground state energy of the system is given by the sum of the
different $\lambda_i$.  Since $\sum_i \delta\lambda_i=0$, we must look
at fluctuations in $\lambda$ to get the fluctuations in the ground state
energy.  From the calculation of appendix A, and from the discussion
of the ``hole" in section XI, we understand that $\lambda$ is being attracted
to this ``hole" which has width $(\frac{k}{V})^{2/3}=k^{-1}$.  If $\lambda$
increases beyond $\sqrt{\frac{1}{\tau}}$ then $e^{-kV\tau\lambda^2}
({\rm det}(H_>+\lambda))^k$ becomes small since the contribution to the
determinant from the semicircle decreases rapidly for 
$\lambda>\sqrt{\frac{1}{\tau}}$.  On the other hand, if $\lambda$ gets
much smaller than the value calculated in appendix A,
$\lambda=-\sqrt{\tau}+\frac{3}{4}\sqrt{\tau}2^{1/3}k^{2/3}$, it will
be difficult to satisfy equation (\ref{poscon}).  So, $\lambda$ varies
over an interval of range $k^{-1}$.  
As $\lambda$ is varied over this interval, one must find how
$e^{-kV\tau\lambda^2}
({\rm det}(H_>+\lambda))^k$ varies; this can be accomplished by
a calculation along the lines of that in appendix A, or by using the
following simple argument.  The attraction between $\lambda$ and the
``hole" is effectively a logarithmic Coulomb attraction between two
charges of opposite sign and magnitude $k$.  The hole has width $k^{-1}$.
A simple calculation on this Coulomb system shows that $\lambda$ can
vary only over a region of width $k^{-2}$, where one factor of $k^{-1}$
arises from the width of the ``hole" and the other factor of $k^{-1}$ arises
from the magnitude of the charges.  So, the fluctuation in the
energy of the system is proportional to $Vk^{-2}=V^{1/5}$ and the
square fluctuation in energy is proportional to $V^{2/5}$.

This may at first seem surprising.  For finite dimensional random systems, the
square fluctuation
in the energy is usually of order $V$ by the central limit theorem;
the total energy is the sum of $V$ different quantities with fluctuations
of order unity.  However, in an infinite range model such as is considered
here, each site is connected to $V$ different sites, and the fluctuations
in the energy for a given site are correspondingly much smaller.  Therefore,
it is possible for the mean square
fluctuation in total energy to be of order $V^{2/5}$
instead of $V$.

In the paramagnetic phase, where the mean-square
fluctuations in $\lambda_i$ are of order $1/V$, the mean-square fluctuation
in total energy is of order unity.

The shift in $\lambda$, leading to $\lambda$ slightly greater than 
$-\sqrt{\frac{1}{\tau}}$ is not surprising.  This implies that the
ground state energy of the large $N$ system is slightly higher than that of
the spherical model.  This is of course not surprising since the
spherical model has additional freedom to choose its ground state, as
the constraint on the sum of the spins from the spherical model is less
constricting than the large $N$ constraint on the length of each individual
spin.

Having obtained these results for the zero temperature problem, we
expect that they will hold true for the spin glass phase.  As discussed
above, the thermodynamic limit and large $\beta$ limit both involve
sending eigenvalues to zero, so the properties of the spin glass
phase should be very similar to the zero temperature problem.
\section{Connection with Efetov Supersymmetry}
We will discuss a slight modification to the previously
constructed supersymmetric formalism.  This modification will make
the formalism very closely
related to the Efetov supersymmetry used in calculations on non-interacting
disordered systems.  It opens the possibility of using field theory
techniques to deal with disordered and glassy $N$ systems in the
large $N$ limit; this might prove much more useful for finite dimensional
systems when the techniques of random matrix theory are not available.
Unfortunately, in the spin glass phase of the
infinite range model it has not yet been possible
to proceed with this modified formalism to the extent that has been done
above with the original formalism.  Still, the technique is interesting in 
itself, and will provide a field-theoretic method of obtaining the properties 
of the paramagnetic phase discussed above.

One complication when trying to apply field theory techniques to equation 
(\ref{rieq}) is the appearance of the Green's
function in the exponential.  This makes it impossible to do what one would
normally like to do, namely integrate over $H$ before integrating over 
any other variables, as it is difficult to integrate over matrices
$H_{ij}$ when the action depends on the inverse of the matrix.  We can
rewrite equation (\ref{rieq}) in a slightly simpler form that avoids
some of these complications.  First we can rewrite it as
\be
\int \limits_{G \ge 0}
dG_{ij} d\lambda_i \, \frac{dc_i}{2\pi} 
\, d\overline a_i \, da_i e^{ic_i(G_{ii}-1)}
e^{\overline a_i G_{ij} a_j G_{ji}}
\delta(G-(H+\lambda)^{-1})
\ee
We can replace the $\delta$-function by $\delta((H+\lambda)G-1) 
{\rm det}(H)^{V}$.  Writing the $\delta$-function by an integral over
a set of auxiliary variables $c_{ij}$ and writing the determinant as
an integral over a set of Grassman variables $\overline\psi_{ij},\psi_{ij}$,
we obtain
\be
\int \limits_{G\ge 0}
dG_{ij} dc_{ij}
d\overline\psi_{ij}d\psi_{ij}
d\lambda_i \, dc_i \, d\overline a_i \, da_i e^{ic_i(G_{ii}-1)}
e^{\overline a_i G_{ij} a_j G_{ji}}
e^{i{\rm Tr}(c(HG-1))}e^{{\rm Tr}(\overline\psi H \psi)}
\ee
This introduces a second supersymmetry into the problem, connecting
the variables $c,G$ with the variables $\overline\psi,\psi$.  This
supersymmetry is closely related to the Efetov supersymmetry used
in calculations for disordered system\cite{efetov}.

In the paramagnetic phase, it is possible to use this formalism very
simply.  It has not yet been possible to obtain anything interesting
in the spin-glass phase using this formalism.  In the paramagnetic phase,
we can integrate over $H$, and then introduce a supermatrix field
$Q$ to decouple the integral, following procedures similar to those
used in the Efetov supersymmetry technique\cite{efetov}.  In the 
paramagnetic phase, a simple saddle point approximation on $Q$ suffices,
since $Q$ has a gap for fluctuations.  At this point, the notation
becomes quite complicated, although the ideas are simple, and not very
different from those used in the Efetov technique used for non-interacting
systems.  We will simply state that such a saddle-point technique reproduces
all the results of the section on the paramagnetic phase of the large $N$
spin glass.

In the spin-glass phase, the fluctuations of the superfield $Q$ become
gapless, and it becomes difficult to proceed with this technique. 
We expect that the bosonic sector of $Q$ may have $k$ eigenvalues separate
out from the others; there will be then $V-k$ remaining bosonic components
and $V$ remaining fermionic ones.  The difference of $k$ between these
may produce the factor of $({\rm det}(H_>+\lambda_i))^k$ that was so crucial
before.  However, it is difficult to use Efetov supersymmetry near the
tail of the spectrum of eigenvalues of a random matrix.  This is the
problem in proceeding further in the formalism of this section.
\section{Troubles with Replica Solutions to the Problem}
We will briefly review the replica solution to the large $N$ spin glass
problem\cite{oldstuff}, indicating a number of mathematical problems.  The
replica solution yields good results in the paramagnetic phase, but does
not correctly address the spin glass phase.  Two important questions
considered in this paper, namely the number of spin components used to
form the spin glass state and the site-to-site or sample-to-sample fluctuations
in $\lambda$, are not even considered within this replica formalism.
However, we can point out a few other mathematical problems.

In the replica technique we compute a partition function such as
\be
Z=\int\limits_{\sum\limits_{\mu} (\phi_i^{\mu,\alpha})^2=N} 
d\phi_i^{\mu,\alpha} e^{\beta H} 
\ee
where $\phi_i^{\mu}$ has now been given an extra index $\alpha$.  This
is a replica index ranging from $1...n$.  In the end, the limit $n=0$ is
taken.

After averaging over $H$, one can then decouple the average with
a replica matrix $Q_{\alpha,\beta,\mu,\nu}$.  The result is
\be
Z=\int\limits_{\sum\limits_{\mu} (\phi_i^{\mu,\alpha})^2=N} 
d\phi_i^{\mu,\alpha} \,dQ_{\alpha,\beta} e^{-V\tau{\rm Tr}(Q^2)}
e^{\phi_i^{\mu,\alpha} Q_{\alpha,\beta,\mu,\nu} \phi_i^{\nu,\beta}}
\ee
Finally, the constraint on the length of the spins is enforced with a
Lagrange multiplier $\lambda_i^{\alpha}$ to obtain
\be
Z=\int\limits_{\sum\limits_{\mu} (\phi_i^{\mu,\alpha})^2=N} 
d\lambda_i^{\alpha}\,
d\phi_i^{\mu,\alpha} \,dQ_{\alpha,\beta} e^{-V\tau{\rm Tr}(Q^2)}
e^{\phi_i^{\mu,\alpha} (Q_{\alpha,\beta,\mu,\nu}+i\lambda_i^{\alpha})
 \phi_i^{\nu,\beta}}
e^{-iN\lambda_i^{\alpha}}
\ee

Taking saddle points in both $Q$ and $\lambda$, one can obtain results
in the paramagnetic phase.  However, there are several problems.  First
of all, the decoupling matrix has a number of components of order $N^2$.
The fact that $Q$ has a gap to fluctuations in the paramagnetic phase
of order $V$ is not very useful in the large $N$ limit, since the number
of components is so great.  It is interesting to note that restricting $N$
to be of order $\sqrt{V}$, remembering the result on the maximum $k$ to
form a ground state, we would find that the matrix Q must still have
of order $V$ different components.  

The fluctuations in $Q$ do have at least one important effect.
If one ignored these fluctuations, one would naively
think that the gap for fluctuations in $\lambda_i$ is of order $N$, since
for fixed $Q$ this is indeed the correct result.  However, integrating over
the $N^2$ different components of $Q$ must reduce the gap for fluctuations
in $\lambda_i$ to be of order $V$, as this is the result found in section VII.

Another problem has to do with the spin glass phase.  In the supersymmetric
formalism, we had to restrict to the sector $H+\lambda_i \ge 0$.  This
might seem to be a problem with the supersymmetric formalism, as this
is a slightly strange restriction to enforce.  However, a similar problem
will arise in the replica formalism.  The procedure used is to integrate
over all $H$ and then decouple this to produce an integral over $Q$.  Then,
an effective action for $Q$ and $\lambda$ is obtained by integrating
over $\phi$.  However, since the fields $\phi$ are bosonic, the integral over
$\phi$ is only well defined if $H+\lambda \ge 0$.  So, the procedure of
integrating over $H$, decoupling, and then integrating over $\phi$, is
ill-defined, unless one also can somehow apply the same restriction
$H+\lambda_i \ge 0$.  One cannot integrate over all $H$ at fixed $\lambda$
and still be assured that $H+\lambda$ will be positive definite.

Further, in the spin glass phase, we saw from the previous sections that
$H+\lambda$ is close to gapless; the gap vanishes as a power law in the
thermodynamic limit.  The replica matrices also will have gapless
excitations.  Considering the calculation of
appendix A, we expect that a detailed consideration of fluctuations
is crucial to getting correct answers here, and cannot easily be obtained
by using replica techniques, or techniques outlined in the section XIII.

Another problem lies in the order of limits.  There are the limits
$n \rightarrow 0$, $N \rightarrow \infty$, and $V \rightarrow \infty$.
Even without the replica limit, there are complications in interchanging
the large $N$ and large $V$ limits, as we have found in this paper.

In conclusion, although the replica technique is very powerful for
certain problems, especially problems with interaction and finite $N$, where
other techniques cannot be used, the old results in the literature for
the large $N$ spin glass using replica techniques should not be trusted
fully.  Although very useful and enlightening in the paramagnetic phase,
in the spin glass phase there will be problems.
\section{Hermitian and Symplectic Systems}
Instead of the system of real Hamiltonians considered in above, we can
consider more general systems.  One possibility it to take the spins $\phi$ to 
be $N$ component complex vectors, instead of $N$ component real vectors.
Then, we can take the matrix $H_{ij}$ to be an arbitrary Hermitian
matrix instead of simply a real symmetric matrix.  So, the
system will be defined by the Hamiltonian
\be
H=\sum\limits_{i,j,\mu}\overline\phi_i^{\mu} \phi_j^{\mu} H_{ij}
\ee
where $i,j$ index the various sites and range from $1$ to $V$, while $\mu$
indexes various components of the spin and ranges from $1$ to $N$.  We
have the constraint that
\be
\sum\limits_{\mu}|\phi_i^{\mu}|^2=N
\ee

We will simply sketch the changes in the Hermitian case from the real
case considered above.  Previously, we derived the bound that
$k(k+1)/2 \leq V$ where $k$ was the number of spin components needed
to form a ground state for a given $H$.  This was based on the fact that
for a real matrix $H+\lambda$, to obtain $k$ zero eigenvalues requires
$k(k+1)/2$ free parameters.  For a general Hermitian matrix $H+\lambda$,
obtaining $k$ zero eigenvalues requires  $k^2$ parameters.  So, we
may obtain the bound that $k^2\leq V$.  Here, $k$ counts the number of
complex spin components.

Similarly, a supersymmetric formalism may be derived for the Hermitian
case.  The formalism of section IX requires works equally well for the
Hermitian case.  The only change required is that we will write
$H+\lambda=U^{\dagger}EU$, instead of
$H+\lambda=O^{\rm T} E O$.  Then, the measure of integration will be different
in the Hermitian case, so that we will write
$\prod\limits_{\mu<\nu} |E_{\mu}-E_{\nu}|^2 |{\rm det}
(H_>+\lambda)|^{2k}$, instead of
$\prod\limits_{\mu<\nu} |E_{\mu}-E_{\nu}| |{\rm det}(H_>+\lambda)|^{k}$.  The
calculation of section XI can still be pursued for the Hermitian case, and
the power law dependences will be the same in the Hermitian case as in the
real case considered above.  

The formalism of section XIII must be slightly changed for the Hermitian
case.  The matrices $G$ and $H$ can now be complex.  To implement
the constraint $\delta(HG-1)$ we will then need to use a complex field $c$.
Using complex $G$ and $c$ amounts to doubling the number of bosonic degrees of
freedom.  We will then need to use two fields $\overline\psi$ and two
fields $\psi$ to get the correct Jacobian.  However, all these changes
are easy to implement.

There is also a symplectic case, in which $H$ is taken from the symplectic
ensemble of matrices.  In the symplectic case, we have the bound
$2k^2-k\leq V$.  Also, the level repulsion factor
in section IX must be replaced with 
$\prod\limits_{\mu<\nu} |E_{\mu}-E_{\nu}|^4 |{\rm det}
(H_>+\lambda)|^{4k}$.  Similarly,
compared to the real case, the symplectic case will require four times as
many fields using the formalism of section XIII.  These changes are also
all easy to implement.
\section{Conclusion}
In conclusion, we have looked at the problem of the large $N$ infinite
range spin glass, and obtained several results on the spin glass phase.
One reason for the interest in the problem is that it is perhaps the
simplest system which combines both disorder and interaction.  It is
simple enough that one feels that there should exist a simple solution,
but, as discussed above, previous attempts to solve this problem do not
fully succeed.  It is perhaps surprising that such a simple problem requires
a technique as complicated as that used here.

This problem exhibits many interesting aspects of disordered systems.  It
has an analogue of Griffiths effects and a non-trivial ground state.
The infinite range spin glass with finite $N$ has been solved by 
replica techniques before.
However, it is nice to have a glassy sytem that can be solved without
using replica techniques, to use as a check.  For certain aspects of
random matrix theory, such as level-level correlation functions, replica
techniques run into troubles, so it is interesting to find a supersymmetric
technique to deal with a glassy system.

The results derived include the scaling of $k$ as $V^{2/5}$; the scaling
of the gap to excitations as $(\frac{k}{V})^{2/3}=V^{-2/5}$; and the magnitude
of fluctuations in $\lambda_i$.  It has
not yet been possible to derive the exact ratio of $k/V^{2/5}$.  It is
interesting that we deal with many quantities which vanish in the thermodynamic
limit in this paper.  The gap, the ratio $k/V$, the correlation functions
in the spin glass phase, and other quantities all vanish in the large $V$
limit, but they do so more slowly than $1/V$, which makes them interesting.

The method introduced in the paper will hopefully be of more general use.
It would be very interesting to try to use this technique to attack problems
with finite range interactions.  Finite dimensional, large $N$ spin
glass systems are one possibility.  Another possibility are finite
dimensional, large $N$ system with disorder, but without frustration.  One
example of these system, the large $N$ dirty boson model\cite{me}, has
been treated with an RG technique.  The supersymmetric technique outlined
here might be useful in that context.  It would also be very interesting
if a method were found for including $1/N$ corrections in this formalism.
For finite $N$ in the large $V$ limit we expect replica symmetry breaking,
so it is not clear what would happen given the first $1/N$ correction to
the large $N$ limit.  The self-consistency equation could have multiple
solutions for finite $N$, and it is not clear what would happen.
\section{Appendix A: Solution of Variational Equation for $\lambda$}
Here we consider a problem in which all $\lambda_i$ are equal to each
other.  We will write $\lambda_i=\lambda$.  The problem is to find
the most likely value of $\lambda$.  The reason for considering
this problem is discussed at the start of section XI.

Equation (\ref{orthoshift}) depends on $\lambda$ in two ways.  First,
there is the factor of $e^{-V \tau {\rm Tr} H^2}$.
Second, there is a factor of $({\rm det}(H_>+\lambda))^k$.  
As discussed at the start of section XI, we must compute
\be
e^{-kV\tau \lambda^2} ({\rm det}(H_>+\lambda))^k
\ee

Let $H_>$ have a density of eigenvalues equal to $V \rho(x)$.
Since $H_>$ has $V-k$ eigenvalues, we will have $\int \rho(x) dx=\frac{V-k}{V}$.
The problem of finding the density of eigenvalues of
$H_>$ and the correct value of $\lambda$ reduces to maximizing the functional
\be
\label{minfunc}
-Vk\tau\lambda^2+
V^2\frac{1}{2}\int dx \, dy\, \rho(x) \rho(y) {\rm log}|x-y|+
V^2\int dx\, \rho(x) (-\tau x^2 +\frac{k}{V} {\rm log}|x+\lambda|)
\ee
where the first integral represent the effects of level repulsion and
the second integral represents the effects of the Gaussian confining
potential and the determinant $({\rm det}(H_>+\lambda))^k$.

First, we will find the density of eigenvalues of $H_>$ for given
$\lambda$.  Then, we will find the optimal value of $\lambda$.
By varying equation (\ref{minfunc}) with respect to $\rho(x)$, and then
differentiating the resulting functional equation with respect to
$x$, we obtain
\be
\label{potfunc}
\int dy \, \rho(y) \rm{log}|x-y|=2\tau x -\frac{k}{V}\frac{1}{x+\lambda}
\ee
Equation (\ref{potfunc}) only holds for $x$ such that $\rho(x)>0$.
We assume that $\rho(x)$ is non-zero for $a<x<b$, and zero for $x<a$ or
$x>b$.

The techniques for solving this equation can be found in the book
\cite{mushkeshvili}.
We simply quote and use the results of that work here.

The function $\rho(x)$ can be obtained from the expression
\be
\label{f1int}
\rho(x)=\frac{1}{\pi^2}\sqrt{(x-a)(b-x)}\int\limits_{a}^{b}
\frac{dy\, 2\tau y-\frac{k}{V}\frac{1}{y+\lambda}}{\sqrt{(a-x)(x-b)}(y-x)}
\ee
This integral can be performed by a series of simple steps: 
write $y=\frac{b-a}{2}z+\frac{a+b}{2}$ so that the integration range
for $z$ extends from $-1$ to $1$.  Then write $z={\rm cos}(\theta)$
to get an integral over $\theta$ from $0$ to $\pi$.  
Finally write $w=e^{i \theta}$, for a complex variable $w$ and extend
the integration range for $\theta$ from $0...\pi$ to $0...2\pi$.
Then, the integral over $w$ is an integral over the unit circle in
the complex plane and can be performed by contour integration.

The result is
\be
\rho(x)=\frac{1}{\pi}\sqrt{(x-a)(b-x)}(2\tau
+\frac{k}{V\sqrt{(a+\lambda)(b+\lambda)}}
\frac{1}{\lambda+x})
\ee
For $k=0$, this reduces to the well known Wigner semi-circle.

There is also a consistency equation that must be satisfied that helps
determine the limits $a$ and $b$.  This equation is
\be
0=\frac{1}{\pi^2}\sqrt{(x-a)(b-x)}\int\limits_{a}^{b}
\frac{dy\, 2\tau y-\frac{k}{V}\frac{1}{y+\lambda}}{\sqrt{(a-x)(x-b)}}
\ee
This integral can be performed by the same contour integration techniques
as the previous one.  The result is
\be
\label{consiseq}
0=\tau (a+b)-\frac{k}{V\sqrt{(a+\lambda)(b+\lambda)}}
\ee
In the case $k=0$, this reduces to the requirement that $a+b=0$.

Finally, we have the constraint that $\int\limits_{a}^{b}\rho(x) \, dx=
\frac{V-k}{V}$.
This integral can also be performed using contour integration.  The
result here is
\be
\label{voleq}
\frac{V-k}{V}=\tau(\frac{b-a}{2})^2-k+\frac{k}{V}\frac{\frac{a+b}{2}+\lambda}
{\sqrt{(b+\lambda)(a+\lambda)}}
\ee
For $k=0$, this equation gives $b-a=\frac{1}{2\sqrt{\tau}}\sqrt{\frac
{V-k}{V}}$.

Putting equations (\ref{consiseq},\ref{voleq}) together, we can obtain
$a$ and $b$.  Let us write
\be
b-a=\frac{2+l}{\sqrt{\tau}}
\ee
\be
\lambda=-a+\frac{\delta}{\sqrt{\tau}}
\ee
We assume that $l$ and $\delta$ are small numbers.  For $k=0$, we
will find that $l$ is of order $k/V$.  We approximate
$\sqrt{(b+\lambda)(a+\lambda)}=\frac{\sqrt{2\delta}}{\tau}$.
Then equation (\ref{consiseq}) becomes 
\be
a+b=\frac{k}{V}\frac{1}{\sqrt{2\delta}}
\ee
Ignoring terms of order $k/V$, or smaller, in equation (\ref{voleq}),
we find 
\be
0=\sqrt{\tau} l+\frac{k}{V}\frac{\tau}{\sqrt{2\delta}}
\ee
Combining these we find that
$a+b=-l/\sqrt{\tau}$, or $b=\frac{1}{\sqrt{\tau}}$ 
up to corrections of order $k/V$.

Finally, we can derive an equation for $\lambda$.  Varying $\lambda$
and requiring that equation (\ref{minfunc}) be stationary gives
\be
\label{lmineq}
2\tau\lambda=\int\limits_{a}^{b} dy\,\rho(x) \frac{1}{y+\lambda}
\ee

It is possible in general to evaluate the following integral
\be
\int\limits_{a}^{b} dy\,\rho(x) \frac{1}{y-x}
\ee
The result is
\be
\sqrt{(\lambda+a)(\lambda+b)}(2\tau\frac{\lambda}{\sqrt{(\lambda+a)
(\lambda+b)}}-2\tau+\frac{k}{V}\frac{1}{x+\lambda}(
\frac{1}{\sqrt{(a-x)(b-x)}}-\frac{1}{\sqrt{(\lambda+a)(\lambda+b)}}))
\ee
Taking a limit as $x$ goes to $\lambda$ in the above equation,
and substituting into equation (\ref{lmineq}), we find
\be
2\tau=\frac{(k/V)(\frac{a+b}{2}+\lambda)}{(\lambda+a)(\lambda+b)^{3/2}}
\ee

Finally, using the consistency and volume equations, as well as the
equation for $\lambda$, we find that $b=\sqrt{\tau}$,
$a=-\frac{1}{\sqrt{\tau}}+\frac{1}{\sqrt{\tau}}2^{1/3}(\frac{k}{V})^{2/3}$, and
$\lambda=\frac{1}{\sqrt{\tau}}-\frac{3}{4}\frac{1}{\sqrt{\tau}}2^{1/3}
(\frac{k}{V})^{2/3}$.
\section{Appendix B: Contribution to $S_k$ of Order $V/\sqrt{k}$}
As discussed in section XI, the calculation of the fluctuations in
$\lambda_i$ and the calculation of the determinant of $M_{ij}$ is
only approximate.  There will be corrections to this calculation which
will produce correction to $S_k$ of order $V/\sqrt{k}$.  

We will not precisely evaluate these terms, but we will show
that they are of order $V/\sqrt{k}$; this is slightly surprising
since there are many other corrections to $S_k$ which are
of order $V/k$.  For example, the calculation of
the integral over $c_i$, so that equation (\ref{con4}), which includes
only Gaussian fluctuations in $c_i$ is not quite right.  If calculated
more carefully, there would be corrections of order $V/k$ from the
cubic and quartic terms in the action $c_i$.

We will look carefully only at the corrections to order $V/\sqrt{k}$, the
correction to the calculation of the determinant of $M_{ij}$.  
The calculation will only be sketched, to save space.  Off-diagonal
terms in $M_{ij}$, after removing terms in $M_{ij}$ of order $\beta^2$ 
as discussed in section XI, are equal to
\be
2\sum\limits_{\mu=1}^{k}\sum\limits_{a=k+1}^{V}
\frac{1}{E_{a}}
\phi^{\mu}_i\phi^{\mu}_j v^{a}_iv^{a}_j
\ee
These terms have mean square $\frac{1}{k}\bigl(G_{ij}^>\bigr)^2$, 
where $G_{ij}^>$
is the Greens function of matrix $H_>+\lambda$.  If one writes the
determinant as $e^{{\rm Tr \, log} M_{ij}}$, and perturbatively
expands the log in the off-diagonal terms, the first correction
one finds is
$e^{\frac{1}{2}\sum\limits_{i,j=1}^{V} \frac{1}{k}\bigl(G_{ij}^>\bigr)^2}$.
This might make one think that these corrections 
will only change $S_k$ by order $V/k$.  However, although in the paramagnetic
phase $\bigl(G_{ij}^>\bigr)^2$ is of order $1/V$ this is not true in
the spin glass phase, and the perturbative expansion diverges and is
therefore not valid.

We can use a trick to get around this.  The desired determinant can be written
\be
\frac{\int d\overline\psi_a^{\mu} d\psi_a^{\mu} 
\prod_i (\sum\limits_{\mu,\nu}
2\overline\psi_a^{\mu} v_i^a \phi_i^{\mu}
\phi_i^{\nu} v_i^a \psi_a^{\nu})
e^{\overline\psi^{\mu}(H_>+\lambda)\psi^{\mu}}
}{({\rm det}(H_>+\lambda))^k}
\ee
Here, $\mu=1...k$ and $a=k+1...V$.  The vectors $v^a$ are the eigenvectors
of $(H_>+\lambda)$.  In the basis of these eigenvectors, $(H_>+\lambda)$
is diagonal.  For simplicity, let us evaluate this with a specific
choice of $\phi_i^{\mu}$.  Let us take $\phi_i^{1}=\beta$ for $i=1...V/k$
and $\phi_i^{1}=0$ for $i=1+V/k...V$.  Also, we will take $\phi_i^2=\beta$
for $i=1+V/k...2V/k$ and zero for all other $i$.  We will follow this
pattern for all other $\mu$, so that for a given $i$, $\phi_i^{\mu}$ is
non-zero only for one value of $\mu$.  
The matrix $H_>+\lambda$ has $V-k$ non-zero eigenvalues, and can be
thought of as a $(V-k)$-by-$(V-k)$ matrix.  Using the particular
choice of $\phi_i^{\mu}$ here, the numerator of the above equation can be
written as the product of $k$ determinants, one for each $\mu$.
Each determinant is the
determinant of a $(V-k-V/k)$-by-$(V-k-V/k)$ random matrix since for the
given choice of $\phi$, for each $\mu$ the factor
$\prod_i (\sum\limits_{\mu,\nu}
2\overline\psi_a^{\mu} v_i^a \phi_i^{\mu}
\phi_i^{\nu} v_i^a \psi_a^{\nu})$ removes $V/k$ different factors of
$\overline\psi^{\mu}, \psi^{\mu}$ from the integral.
In fact, this result is not specific to this choice of $\phi_i^{\mu}$, but
is easiest to obtain with this particular choice.

Now, if we evaluate the product of the determinants of the $k$ different
$V-k-V/k$-by-$V-k-V/k$ random matrices, and divide by the $k$-th power
of the determinant of $H_>+\lambda$, we find that the result is
$(2\sqrt{\tau})^V$, times some corrections which change $S_k$ by terms of
order $V/\sqrt{k}$.  The calculation of the ratio of these determinants
is elementary and will not be done here.  It is simply a matter of using
the Wigner semicircle distribution for the eigenvalues of the different
matrices, and evaluating the determinant by taking a product of eigenvalues.

Similarly, there are corrections to the fluctuations in $\lambda_i$,
involving the contribution to these fluctuations resulting from
$({\rm det}(H_>+\lambda_i))^k$.  Looking at a perturbative expansion
of the determinant similar to that performed above, one would again 
think that the corrections to
the fluctuations change $S_k$ by order $V/k$, but again the perturbation
expansion involves $(G_{ij})^2$ and is invalidated in the spin glass phase.
One again finds corrections to $S_k$ of order $V/\sqrt{k}$.  

Finally, the corrections to $S_k$ discussed here depend on the average
value of $\lambda_i$.  This can lead to a shift in $\lambda$ away from
the value calculated in appendix A.  These corrections can adjust the
value of the gap, $\lambda+a$, but for $k\propto V^{2/5}$
they will not alter the scaling of the gap with $k$; if we calculate an
effective action for $\lambda$ combining the corrections discussed here
with the results from appendix A, all the terms will be of the same
magnitude and $\lambda+a$ will still scale as $(\frac{k}{V})^{2/3}$.

\end{document}